\documentclass[twocolumn]{aastex631}

\usepackage{hyperref}
\usepackage{amsmath}
\usepackage{multirow}
\begin{document}

\title{Investigating silicate, carbon, and water in the diffuse interstellar medium:\\ the first shots from WISCI}

\correspondingauthor{S.T. Zeegers}
\email{sascha.zeegers@esa.int}

\author[0000-0002-8163-8852]{S.T. Zeegers}
\affiliation{European Space Agency (ESA), European Research and Technology Centre (ESTEC), Keplerlaan 1, 2201 AZ Noordwijk, The Netherlands}

\author[0000-0001-6208-1801]{Jonathan P. Marshall}
\affiliation{Institute of Astronomy and Astrophysics, Academia Sinica, 11F of AS/NTU Astronomy-Mathematics Building, No.1, Sec. 4, Roosevelt Road, Taipei 106319, Taiwan}

\author[0000-0001-5340-6774]{Karl D.\ Gordon}
\affiliation{Space Telescope Science Institute, 3700 San Martin Drive, Baltimore, MD, 21218, USA}
\affiliation{Sterrenkundig Observatorium, Universiteit Gent, Krijgslaan 281 S9, B-9000 Gent, Belgium}

\author{Karl A. Misselt}
\affiliation{Steward Observatory, University of Arizona, Tucson, AZ 85721, USA}

\author[0000-0002-6717-1977]{G.P.P.L. Otten}
\affiliation{Institute of Astronomy and Astrophysics, Academia Sinica, 11F of AS/NTU Astronomy-Mathematics Building, No.1, Sec. 4, Roosevelt Rd, Taipei 106319, Taiwan}

\author[0000-0003-4757-2500]{Jeroen Bouwman}
\affiliation{Max-Planck-Institut f\"ur Astronomie, K\"onigstuhl 17, D-69117 Heidelberg, Germany}

\author[0000-0003-2029-1549]{Jean Chiar}
\affiliation{Physical Science Department, Diablo Valley College, 321 Golf Club Road, Pleasant Hill, CA 94523, USA}

\author[0000-0001-9462-5543]{Marjorie Decleir}
\affiliation{European Space Agency (ESA), ESA Office, Space Telescope Science Institute, 3700 San Martin Drive, Baltimore, MD 21218, USA}

\author[0000-0002-9583-5216]{Thavisha Dharmawardena}
\affiliation{Center for Computational Astrophysics, Flatiron Institute, 162 5th Ave, New York, NY 10010, USA}

\author[0000-0003-2743-8240]{F. Kemper}
\affiliation{Institut de Ciències de l'Espai (ICE, CSIC), Can Magrans, s/n, E-08193 Cerdanyola del Vallès, Barcelona, Spain}
\affiliation{Institució Catalana de Recerca i Estudis Avançats (ICREA), Pg. Lluís Companys 23, E-08010 Barcelona, Spain}
\affiliation{Institut d'Estudis Espacials de Catalunya (IEEC), E-08860 Castelldefels, Barcelona, Spain}

\author[0000-0002-1119-642X]{Aigen Li}
\affiliation{Department of Physics and Astronomy, 223 Physics Bldg., University of Missouri, Columbia, MO 65211, USA}

\author[0000-0002-0554-1151]{Mayank Narang}
\affiliation{Institute of Astronomy and Astrophysics, Academia Sinica, 11F of AS/NTU Astronomy-Mathematics Building, No.1, Sec. 4, Roosevelt Rd, Taipei 106319, Taiwan}

\author[0000-0002-3699-7477]{Alexey Potapov}
\affiliation{Analytical Mineralogy Group, Institute of Geosciences, Friedrich Schiller University Jena, CEEC II, Lessingstr. 14, 07743 Jena, Germany}

\author[0000-0002-9497-8856]{Manoj Puravankara}
\affiliation{Tata Institute of Fundamental Research, Mumbai, Maharashtra, India}

\author[0000-0002-1161-3756]{Peter Scicluna}
\affiliation{Centre for Astrophysics Research, Department of Physics, Astronomy and Mathematics, College Lane Campus, University of Hertfordshire, Hatfield AL10 9AB, UK}

\author[0000-0002-9497-8856]{Himanshu Tyagi}
\affiliation{Tata Institute of Fundamental Research, Mumbai, Maharashtra, India}

\author[0000-0003-3769-8812]{Eleonora Zari}
\affiliation{Max-Planck-Institut f\"ur Astronomie, K\"onigstuhl 17, D-69117 Heidelberg, Germany}
\affiliation{Dipartimento di Fisica e Astronomia, Universit{\`a} degli Studi di
Firenze, Via G. Sansone 1, I-50019, Sesto F.no (Firenze), Italy}

\author[0009-0005-1543-6944]{ChuanYu Wei}
\affiliation{Anton Pannekoek Institute for Astronomy, Universiteit van Amsterdam, Science Park 904, 1098 XH Amsterdam, The Netherlands}

\author[0000-0001-8025-8981]{Lex Kaper}
\affiliation{Anton Pannekoek Institute for Astronomy, Universiteit van Amsterdam, Science Park 904, 1098 XH Amsterdam, The Netherlands}

\author[0000-0003-3670-3181]{Frank Backs}
\affiliation{Institute of Astronomy, KU Leuven, Celestijnenlaan 200D, 3001 Leuven, Belgium}
\affiliation{Anton Pannekoek Institute for Astronomy, Universiteit van Amsterdam, Science Park 904, 1098 XH Amsterdam, The Netherlands}

\author[0000-0002-7037-0475]{Stefan T. Bromley}
\affiliation{Departament de Ci\`encia de Materials i Qu\'imica F\'isica \& Institut de Qu\'imica Te\`orica i Computacional (IQTCUB), Universitat de Barcelona, c/Mart\'i i Franqu\`es 1-11, E-08028 Barcelona, Spain}
\affiliation{Institució Catalana de Recerca i Estudis Avançats (ICREA), Pg. Lluís Companys 23, E-08010 Barcelona, Spain}
\affiliation{Institut d'Estudis Espacials de Catalunya (IEEC), E-08860 Castelldefels, Barcelona, Spain}

\author[0000-0002-1437-4463]{Laurie Chu}
\affiliation{NASA Ames Research Center, Moffett Field, CA 94035, USA}

\author[0000-0001-8470-749X]{Elisa Costantini}
\affiliation{SRON Netherlands Institute for Space Research, Niels Bohrweg 4, 2333 CA Leiden, The Netherlands}

\author[0000-0003-2824-3875]{T. R. Geballe}
\affiliation{Gemini Observatory/NSF’s NOIRLab, 670 N. A’ohoku Place, Hilo, Hawai’i, 96720, USA}

\author[0000-0003-1665-5709]{Joel D. Green}
\affiliation{Space Telescope Science Institute, 3700 San Martin Drive, Baltimore, MD, 21218, USA}

\author[0000-0002-4634-5966]{Chamani Gunasekera}
\affiliation{Space Telescope Science Institute, 3700 San Martin Drive, Baltimore, MD, 21218, USA}

\author[0000-0002-2449-0214]{Burcu G\"unay}
\affiliation{Armagh Observatory and Planetarium, Armagh, NI, UK}

\author[0000-0002-1493-300X]{Thomas Henning}
\affiliation{Max-Planck-Institut f\"ur Astronomie, K\"onigstuhl 17, D-69117 Heidelberg, Germany}

\author[0000-0003-4870-5547]{Olivia Jones}
\affiliation{UK Astronomy Technology Centre, Royal Observatory, Blackford Hill, Edinburgh, EH9 3HJ, UK}

\author[0000-0001-7154-8450]{Joan Mari\~noso Guiu}
\affiliation{Departament de Ci\`encia de Materials i Qu\'imica F\'isica \& Institut de Qu\'imica Te\`orica i Computacional (IQTCUB), Universitat de Barcelona, c/Mart\'i i Franqu\`es 1-11, E-08028 Barcelona, Spain}

\author[0000-0003-1878-327X]{Melissa McClure}
\affiliation{Leiden Observatory, Leiden University, Einsteinweg 55, 2333 CC Leiden, The Netherlands}

\author[0000-0001-8102-2903]{Yvonne J. Pendleton}
\affiliation{University of Central Florida, Dept. of Physics, Orlando, FL, 32816, USA}

\author[0000-0001-6326-7069]{Julia C. Roman-Duval}
\affiliation{Space Telescope Science Institute, 3700 San Martin Drive, Baltimore, MD, 21218, USA}

\author[0000-0003-0642-8107]{Tomer Shenar}
\affiliation{The School of Physics and Astronomy, Tel Aviv University, Tel Aviv 6997801 Israel}

\author[0000-0003-0306-0028]{Alexander G.G.M. Tielens}
\affiliation{Leiden Observatory, Leiden University, Einsteinweg 55, 2333 CC Leiden, The Netherlands}

\author[0000-0002-5462-9387]{L. B. F. M. Waters}
\affiliation{Department of Astrophysics/IMAPP, Radboud University, PO Box 9010, 6500 GL, Nijmegen, The Netherlands}
\affiliation{SRON Netherlands Institute for Space Research, Niels Bohrweg 4, 2333 CA Leiden, The Netherlands}

%\collaboration{40}{(WISCI collaboration)}

\begin{abstract}

The dusty interstellar medium (ISM) of the Milky Way is distributed in a complex, cloudy structure. 
It is fundamental to the radiation balance within the Milky Way, provides a
reaction surface to form complex molecules, and is the feedstock for future generations of stars and planets. 
The life cycle of interstellar dust is not completely understood, and neither are its structure nor composition. 
The abundance, composition, and structure of dust in the diffuse ISM can be determined by combining infrared, optical and ultraviolet spectroscopy.
\textit{JWST} enables measurement of the faint absorption of ISM dust grains against bright stars at kiloparsec distances across the infrared spectrum. 
Here we present an overview of the project `Webb Investigation of Silicates, Carbons, and Ices' (WISCI) along with interpretation of two targets, GSC 08152-02121 and CPD-59 5831. 
Observations of 12 WISCI target stars were taken by \textit{JWST}, \textit{the Hubble Space Telescope}, Himalayan Chandra Telescope, and the Very Large Telescope. We use these to characterize the targets' spectral types and calculate their line-of-sight extinction parameters, $A_{\rm V}$ and $R_{\rm V}$.
We find absorption in the \textit{JWST} spectra of GSC 08152-02121, and CPD-59 5831 associated with carbonaceous dust around 3.4 and 6.2~\micron\ and amorphous silicates at 9.7~\micron. In GSC 08152-02121 we also find indications of absorption by trapped water around 3~\micron\. This first look from WISCI demonstrates the line-of-sight variability within the sample, and the program's potential to identify and correlate features across ultraviolet to mid-infrared wavelengths.
\end{abstract}

\keywords{Interstellar dust (836) --- Infrared astronomy (786) --- Interstellar medium (847) --- Milky Way Galaxy (1054)}

\section{Introduction} \label{sec:intro}

There are many topics in astronomy that are influenced in some way by the existence of interstellar dust \citep{1995Dorschner}. Dust plays an essential role in the evolution of galaxies through its moderation of the interstellar radiation field. At the largest scales, dust dictates the total energy balance of a galaxy by absorbing and re-radiating stellar light. At the smallest scales, submicron-sized dust grains drive the chemical evolution of the interstellar medium (ISM) by providing the reaction surfaces necessary for gas and solid-state molecules to form.

Dust grains may be both created and destroyed within the ISM~\citep{2011Jones}. During their existence, dust grains cycle through clouds in the ISM with a range of densities, from diffuse clouds to dense molecular clouds. The diffuse ISM sets the initial conditions for material that feeds into dense molecular clouds. These dense clouds may eventually collapse to form protostars, and subsequent planetary systems. The diffuse ISM is thus the starting point for star and planet formation. 

Dust extinguishes the starlight that passes through it, and is more efficient at shorter wavelengths \citep[e.g.,][]{1979SavageMathis}. 
On average, visual extinction ($A_\mathrm{V}$) is about two magnitudes per kiloparsec toward stars in the Milky Way \citep{2005Tielens}. 
The \textit{Gaia} survey opened a new window on the distribution of dust clouds in the nearby ($<$~3 kpc) ISM and detailed dust maps of this environment are now available~\citep{2022Lallement, Dharmawardena24}.   
Individual diffuse dust clouds typically provide an extinction of $A_\mathrm{V} < 1~$mag. Along a single line of sight the extinction may be the aggregate effect of several clouds with individual $A_\mathrm{V}$s of 0.2 - 0.6 mag~\citep{Bohlin78}. 
Although these variations are small, the conditions in these clouds to which the dust grains are exposed may be sufficiently different to change the dust grain properties.

Abundant elements, such as carbon, oxygen, silicon, iron, and magnesium are depleted from the cold gas phase and are thought to be locked up in dust~\citep{Savage1996, Jenkins2009}. There is a general agreement that dust in the diffuse ISM consists mostly of silicate and carbonaceous material \citep{1977Mathis}. Important diagnostic information on the diffuse ISM dust is revealed in spectroscopic observations through the variety, shape, and strength of extinction features present from far-ultraviolet to mid-infrared wavelengths. It is as yet not clear if the silicate and carbonaceous dust components exist separately~\citep[e.g.,][]{Draine03} or as a composite material~\citep[e.g.,][]{Hensley2023,Ysard2024}. As a further complication, variations may exist in the dust content and structure in different environments, because the densities of diffuse clouds vary between regions, as we know for the Milky Way \citep{2005Tielens}.

%Mention sightline dependence of extinction laws plus models by Fitzpatrick + Hensley/Draine here?
In the ultraviolet, the far-ultraviolet rise and the $2175$\,\AA{} feature can be observed \citep{1965Stecher, Fitzpatrick86}. These features are associated with the presence of small graphitic carbonaceous grains~\citep{Schnaiter1998} or medium-sized PAHs~\citep{Steglich2010,Fitzpatrick86}. The broad intermediate-scale structures detected in the optical may have the same origin \citep{Massa20}. The optical and near-infrared spectra also harbor many broad, low-amplitude features called diffuse interstellar bands~\citep[DIBs,][]{2017Cox}. These have been linked to small carbonaceous grains in two cases \citep{Campbell15}, but the origin of the majority remains mysterious \citep{1995EhrenfreundFoing}. All these features can be observed at low line-of-sight extinction levels; e.g. the $2175$\,\AA{} feature can still be observed at an $A_\mathrm{V}$ of 0.5 mag~\citep[e.g.,][]{Xiang17,Massa2022}. DIBs can also be detected at very low extinction; for example, the $8620\,$\AA{} DIB feature is explored down to $A_\mathrm{V}$ of 0.12 mag~\citep{Lallement2024}. 

At NIR wavelengths vibrational modes of carbonaceous dust particles can be observed \citep{1998HenningSalama}. 
Interstellar carbonaceous dust contains aliphatic hydrocarbons (single bonded carbons), olefinic hydrocarbons (double bonded carbons), and aromatic hydrocarbons (conjugated carbon rings). 
The 3.2 - 3.65~\micron\ spectral region contains both aromatic and aliphatic (and, likely, olefinic) C-H absorption bands, while the 5.6 - 6.5~\micron\ region contains signatures of olefinic and aromatic C=C stretch absorption bands. Additional weaker aliphatic features may be found at 6.85 and 7.25~\micron. 
Before \textit{JWST} \citep{JWST}, interstellar dust features in the 3.2 - 3.65~\micron\ region were observed with the NASA Infrared telescope Facility~\citep[IRTF,][]{Sandford1991} and the United Kingdom Infrared Telescope  (UKIRT) Facility, among others, and the features in the 5 - 8~\micron\ with the \textit{Kuiper Airborne Observatory}~\citep{Tielens1996} and \textit{Infrared Space Observatory} \citep[ISO,][]{1995Kessler} down to an $A_\mathrm{V}$ of $\sim 5-7$ mag~\citep{Pendleton94,Schutte1998,Rawlings2003}, indicating that hydrocarbons exist at these line-of-sight extinctions.

Also, at NIR wavelengths an absorption feature from solid-state water in dense molecular clouds is found at 3.0~\micron. The dividing line between the dense and diffuse ISM may be the presence or absence of ices. % Add that water ice is often the first to form?
In dense molecular clouds, water ice absorption is typically found when the density increases and ice mantles form on dust grains. In the Rho Ophiuchus dense molecular cloud, \citet{1978Harris} found that ice formed along sight lines with extinction $A_{\rm V} \geq 10 - 15$~mag. Water ice has been subsequently detected at significantly lower values of visual extinction and is a primary component of dense molecular clouds. At the edges of molecular and star forming clouds water ice has been detected down to an $A_\mathrm{V}$ of 3~\citep{Boogert2013,Whittet2001}. 
In general, water ice is not seen at lower $A_\mathrm{V}$ levels in dense clouds or in the diffuse interstellar medium. However, a very low level of water absorption has been identified toward Cyg OB 2 No. 12 \citep[$A_{\rm V} = 10$ mag,][]{Potapov21}, and a recent tentative claim has been made toward HD~73882 \citep[$A_{\rm V}$ = 2.46 mag,][]{Decleir2025}. This water is thought to exist in the form of water molecules trapped (strongly bonded) by silicates \citep{Potapov21}, not as water ice itself.% of the solar abundance of oxygen}

At MIR wavelengths, the 9.7~\micron\ (Si-O stretch) and 18~\micron\ (O-Si-O bend) features of amorphous silicates dominate the absorption spectrum \citep[e.g.,][]{Roche1984,Roche1985,Bowey1998, 2004Kemper, Chiar2006, VanBreemen2011}. Using the  \textit{Spitzer} Space Telescope \citep{2004Werner} spectroscopy~\citet{Fogerty16} and \citet{Gordon2021} observed these features down to an $A_\mathrm{V}$ $\sim1-2$ mag. The shapes of these silicate features are diagnostic of the composition and structure of the dust \citep{2010Henning}. Several diffuse ISM studies have found a mixture of magnesium-rich silicate dust with an amorphous olivine and pyroxene stoichiometry \citep{Chiar2006, 2007Min,Fogerty16}. The diffuse interstellar silicates are overwhelmingly amorphous \citep[crystallinity of 1\%-2\%, ][]{2004Kemper,2005Kemper,2007Li,2007Min}. The sharply peaked features of crystalline grains are absent in the smooth silicate features; although for some diffuse sight lines, a feature at 11.1~\micron\ may indicate the presence of crystalline forsterite~\citep{DoDuy2020,Gordon2021,Shao2024}. Furthermore, the shape of the features may be slightly altered by the presence of nanosilicates~\citep{2023Zeegers}.
 
All of these dust features hold information about the dust properties, which in turn illuminate the evolution of dust grains. The size distribution, chemical composition, and lattice structures of dust grains in the ISM reflect the impact of the conditions to which they have been exposed in the diffuse ISM.
For example, the low degree of crystallinity in silicates may reflect processing in the harsh environment of the diffuse ISM, since silicate dust freshly produced by asymptotic giant branch stars can have high crystalline fractions~\citep{Sylvester1999,2024Cox}. Amorphizing processes may include cosmic-ray hits~\citep{Bringa07}, grain–grain collisions, or atomic impacts in
shocks. These processes may create a large number of nano-sized grains. On the other hand, small amorphous dust grains may also accrete atomic or molecular species in the ISM~\citep[e.g.,][]{Zhukovska2008}. The chemical composition of the dust grains can help distinguish between these processes \citep{2007Min}. 
Additionally, the relative abundances of carbon groups depend on the interaction of the grains with UV light and atomic hydrogen. UV radiation converts aliphatic and olefinic materials into aromatic materials. Hydrogen converts aromatic materials into olefinic and then aliphatic materials~\citep{Mennella_02_CH, Dartois05}. Furthermore, hydrogen can also be removed by collisions with cosmic rays~\citep[e.g.,][]{Gerber2025,Mennella2003}.
\citet{Sorrell89, Sorrell90} and \citet{Hecht91} proposed that hydrogenated amorphous carbon (HAC) grains or HAC coatings on silicate grains entering the diffuse ISM lose most of their hydrogen and become aromatic, thus increasing the strength of the 2175 \AA\ feature. Upon entering denser media, the grains would react with hydrogen, which disrupts the aromatic structures and weakens the 2175 \AA\ feature~\citep{Sorrell91}.  

Acquisition of detailed spectroscopic observations of the diffuse ISM is challenging. 
With the exception of strong silicate features, the 2175\,\AA\, feature, and strong DIBs, many of the absorption features produced by dust (e.g., hydrocarbons and nanosilicates) in the diffuse ISM are shallow and thus faint~\citep[e.g.,][]{Rawlings2003,Gordon2021,2023Zeegers}. 
This requires high signal-to-noise ratios to detect and characterize them, making such features extremely difficult to study at low extinction. The most detailed studies of the faint diffuse ISM features have been obtained in the direction of the Galactic center or more nearby yet highly extinguished sources (e.g.,  Cyg OB 2 No.~12 with $A_\mathrm{V} = 10~$mag), where sufficient dust column densities are present to produce strong absorption bands \citep{Pendleton94, Chiar13, Whittet97}. 
However, these sight lines may not be representative of the dust populations found in the general diffuse ISM, since the observations may trace denser environments and show the presence of ices \citep[e.g.,][]{Sandford1995}. In case of the extremely luminous blue hypergiant Cyg OB 2 No.~12 the high extinction ($A_\mathrm{V} = 10~$mag) is due in part to clouds local to the star and thus not necessarily representative of the diffuse ISM~\citep{Maryeva2016}. 
This confusion can be avoided by observing the local diffuse ISM on sight lines to nearby stars. Bright OB stars have properties appropriate for this task; they are luminous from UV to IR wavelengths, relatively common, contain relatively few lines, and usually lack complex line profiles at optical and NIR wavelengths that would interfere with the detection and interpretation of dust absorption features \citep[e.g.,][]{1965Stecher, Shao2018, Hensley2020, Decleir22, 2023Gordon}.

At lower extinction, the detection limits of \textit{Spitzer} and \textit{ISO} start to play a role when interpreting the spectra~\citep{Rawlings2003, Hensley2020, Gordon2021}.  
The sensitivity and spectral resolution of \textit{JWST} greatly enhances studies of interstellar dust properties at infrared wavelengths in the local diffuse ISM ($<~3$~kpc) providing spectra in unprecedented detail for a population of local OB stars. 
To better infer the properties and composition of dust grains within the diffuse ISM, we instigated the program ``Webb Investigation of Silicates, Carbons, and Ices'' (WISCI program).
This survey targets lines-of-sight through the diffuse ISM toward 12 O and B type stars. These bright stars are observed over a broad wavelength range (UV/optical – MIR), encompassing all important diffuse ISM dust features, including detailed observations of the diagnostic faint features mentioned above. The data set allows for the consistent determination of dust feature properties and comparison of their relative strengths along these lines of sight as the same dust column is responsible for all features in the observed spectra.
The sources were selected based on their line-of-sight extinction and their lack of discernible infrared excess from strong winds or emission lines, thus providing a cleaner background upon which dust features would be imprinted.
Spectra spanning optical to mid-infrared wavelengths (0.4 to 28~\micron) have been obtained for each star, and have been complemented with ultraviolet spectroscopy (down to 0.1~\micron) for seven out of the 12 stars, with the remaining five stars being too faint to be detected. 
The WISCI program falls in between two other dust-focused \textit{JWST} ISM projects, namely ``Ice Age'' \citep{2023McClure} and ``Measuring Extinction and Abundances of Dust'' \citep[MEAD;][]{Decleir2025}. Ice Age ($A_{\rm V} \simeq 60-100~$mag) focuses on the dense ISM where ices are known to be present, whereas MEAD ($A_{\rm V} < 3~$mag) targets more diffuse sight lines than WISCI. 

In this paper we present the initial results from WISCI. We describe the spectral typing and characterize the general dust extinction for the full sample.
For two sources, GSC 08152-02121 (2MASS 08574757$-$4609145) and \mbox{CPD-59 5831} (2MASS J15095841$-$5958463), we identify absorption features from both silicate and carbonaceous dust. These sources were selected to highlight all the observable dust features in the spectra. The sources contrast in brightness and line-of-sight extinction to highlight the full potential of the program. 
% Check fluxes 

The remainder of the paper is laid out as follows. In Section \ref{sec:wisci} we summarize the motivation and sample selection for WISCI, with an overview of the program's observations given in Section \ref{ssec:obs}. We then present results of stellar spectral typing (Section \ref{sec:star_properties}), extinction fitting (Section \ref{sec:extinction}), and the characterization of observed dust features (Section \ref{sec:dust_properties}). We contextualize these results in relation to existing measurements of the composition and size of dust grains within the diffuse ISM in Section \ref{sec:dis}. Finally, in Section \ref{sec:con}, we present our conclusions and give an outlook for the future study of the full dataset.

\section{The Webb Investigation of Silicates, Carbons, and Ices (WISCI) project}
\label{sec:wisci}

Here we summarize the motivation, preparation, and observations taken in support of the WISCI project. In brief, we first determined the feasibility of detection for faint dust absorption features by \textit{JWST} toward a representative OB star, taking into account ISM chemistry, grain size, and stellar luminosity; see Section \ref{sec:feasibility}. Twelve sight lines were selected for the WISCI project based on the brightness and spectral types of the background stars, and the interstellar extinction inferred from near-infrared colors, explained in detail in Section~\ref{sec:sourceselec}. 

\subsection{Feasibility of detecting ISM dust features in the nearby diffuse ISM}
\label{sec:feasibility}

% First info on the feasibility study 
In preparation for this project and before the selection of background sources, we calculated the amount of extinction necessary for the detection of diffuse dust features. 
Stellar atmospheric models of potential background targets were generated with variations in temperature and radius appropriate to main sequence O and B stars \citep{2012Pecaut,2013Pecaut} and for distances up to 3~kpc using Tlusty, BOSZ and BT NEXT-GEN models \citep{1995Hubeny,2003Lanz,2007Lanz,2011Allard,2024Meszaros}. 
The model spectra were then extinguished using the~\citet{1999Fitzpatrick} dust extinction function with $R_\mathrm{V} = 3.1$, implemented in the Python package {\sc extinction} \citep{2021Barbary}. This smooth extinction function does not contain silicate or carbonaceous dust features in the infrared, allowing for testing different models of these features. 
The magnitudes of extinction were used to obtain dust abundances. Firstly, the hydrogen column densities along the lines of sight were calculated using the standard relation  
$N_\mathrm{H}$ /$A_\mathrm{V} = 1.9 \times 10^{21}\,\mathrm{cm}^{-2}~\mathrm{mag}^{-1}$ \citep{1973Reina,Bohlin78}. Secondly, we used abundances from \citet{Lodders2009} to obtain the expected column densities of silicon and carbon. Finally, the atomic dust abundance of carbon and silicon were obtained from dust depletion values of these elements taken from \citet{Wilms2000}. These abundances were then used in conjunction with the expected grain optical properties to predict the optical depths of their respective features.

\textbf{Silicates}: The silicate dust optical constants used in our modeling were taken from \cite{1994Jaeger}. From them we derived extinction efficiencies, assuming the dust grain shape distribution is a continuous distribution of ellipsoids~\citep[CDE,][]{BohrenHuffman83} with an average size of 0.1~\micron. The optical depths of the silicate features were calculated using the following equation: 
$\tau_{\mathrm{sil}} = Q_{\rm ext} a N_{\rm d}$, where $Q_{\rm ext}$ is the extinction efficiency, $a$ is the radius of the dust grain and $N_d$ is the number of grains along the line of sight. 
The absorption depth of the simulated 10 $\micron$\ silicate feature is expected to be between 10 and 25\% relative to the stellar continuum for these sight lines.
The required sensitivity for the silicate band is dictated by the detection of the crystalline features and the possible presence of nanosilicates in the spectrum \citep{2023Zeegers}. We aim to detect feature depths between 1 and 0.5\% relative to the stellar continuum, this is equivalent to a sensitivity to crystallinity at the less than the 1\% level.
With the same requirements, we are able to measure variations in iron content and stoichiometry of the silicates. The 10 and 20~\micron\ bands are both sensitive to iron content. Changes in the stoichiometry (i.e. ratio of olivines and pyroxenes) alter the spectral profile of these bands. 

\textbf{Carbonaceous dust}: The optical depths of the the aromatic ($3.28, 6.2$~\micron) and aliphatic ($3.4$~\micron) features were calculated using the absorption strengths taken from \citet{Chiar13}. The intrinsic absorption strength of the bands have been determined from laboratory studies of hydrocarbon films~\citep[][ and references therein]{Chiar13}. The absorption strengths of the 6.85 and 7.25 \micron\, aliphatic features were taken from ~\citet[][ and references therein]{Chiar00}.  
For O and B stars at distances of 2 to 3 kpc, we expected feature depths from 1 to 2\% for the strongest carbonaceous dust feature. 

\textbf{Water}: We used the results from~\citet{Potapov21} to analyze the possible presence of embedded solid-state water in silicate grains along the line of sight. \citet{Potapov21} derived an $\mathrm{H}_{2}\mathrm{O}$ column density N($\mathrm{H}_{2}\mathrm{O}$) of $3\times10^{16}~\mathrm{cm}^{-2}$ between 2.75 and 3.2~\micron\ along the line of sight toward Cyg OB 2 No. 12 with an optical depth of $\tau=0.025$.  The column density of hydrogen was estimated to be $N_\mathrm{H}= 4\times10^{21}~\mathrm{cm}^{-2}$, which resulted in a $\mathrm{N}(\mathrm{H}_2\mathrm{O})$/$\mathrm{N(H)}\sim $10 ppm. Since this is the only source with a diffuse sightline for which this feature has been modeled, we use it as a benchmark. Assuming that the features scales with the hydrogen column density, which in turn depends on extinction, we conservatively expect feature depths of $0.4 - 0.8 \%$ relative to the stellar continua of our sources. 

Following this analysis, the range of extinctions probed by WISCI, from $A_\mathrm{V}=$ 4 to 8 mag, was selected to avoid sight lines toward regions of dense ISM, defined by the presence of ice features, but with enough dust extinction to produce detectable features based on the assumed ISM dust composition and atomic budget.
The simulation tool developed to test the feasibility of the project can be found on the WISCI GitHub 
page\footnote{\href{https://gitfront.io/r/SZeegers/upp1v9WSCAzk/proposal-simulation-tool/}{https://github.com/WISCI/proposal\_simulation\_tool}}.

\subsection{Selection of background sources}
\label{sec:sourceselec}

O and B type stars can have strong stellar winds and feature emission lines in their spectra in the infrared, which may hamper the detection of dust features and makes them challenging to model. This required a careful selection of possible targets. Our selection relied on a combination of criteria, which we describe in detail below.

\begin{figure}[!ht]
\includegraphics[width=0.5\textwidth]{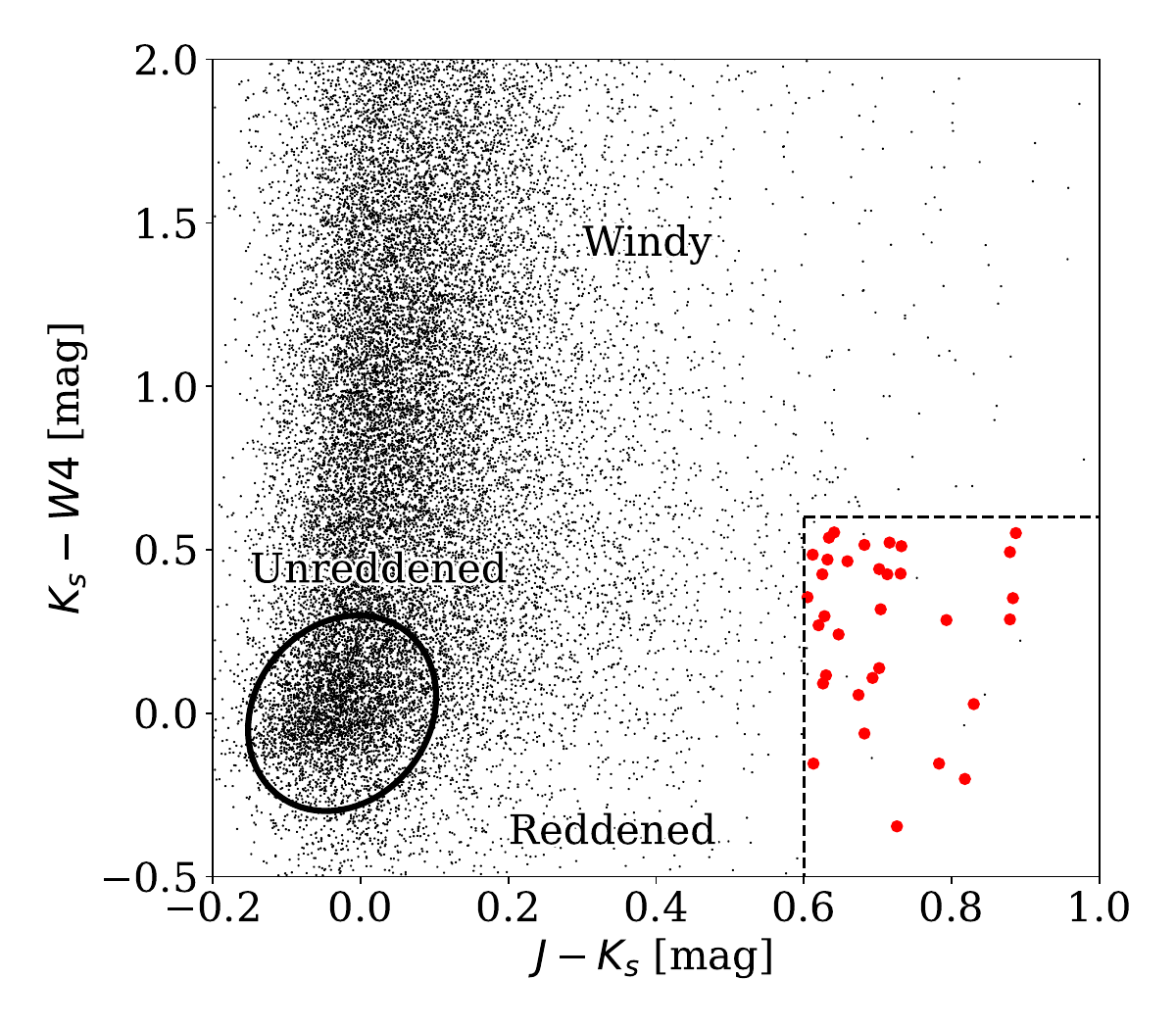}
\caption{$J - K_S$ vs. $K_S - W4$ color-color diagram of spectroscopically confirmed
O and B-type stars \citep[black dots based on the selection from][]{Zari2021} and of those following the criteria in Eq. \ref{eq:one} (red dots). Labels show the approximate regions of different kinds of stars within this color-color parameter space, with the black ellipse encompassing `unreddened' stars.
\label{fig:sourceselection}}
\end{figure}

\begin{enumerate}
     \item {Selection of candidate O- and B-type stars: We used a combination of photometric and astrometric criteria, based on \textit{Gaia} DR2 \citep{2018Gaia} and 2MASS \citep{2006Skrutskie} astrometry and photometry to select bright stars, whose colors are consistent with (reddened) O- and B-type stars\footnote{We note that at the time of the proposal submission, \textit{Gaia} EDR3 \citep{2021Gaia} had not yet been released, while \citet{Zari2021} made use of the \textit{Gaia} EDR3 data. For the purposes of this study, the use of \textit{Gaia} DR2 or EDR3 parallaxes did not make any significant difference.}. In this study we cross-matched the \citet{Zari2021} catalogue with the AllWISE catalogue \citep{2010Wright} (see next item) and with the Simbad astronomical database\footnote{{\url{https://simbad.cds.unistra.fr/simbad/}}} \citep{2000Wenger}. The cross-match with Simbad was motivated by the need to restrict our sample to spectroscopically confirmed O- and B-type stars. The spectral types obtained through Simbad were checked with literature references.} 
     \item {Identification of reddened sources: To select reddened sources without strong infrared emission, we compared the near-IR \citep[2MASS;][]{2006Skrutskie} and mid-IR \citep[\textit{WISE};][]{2010Wright} colors of the sources selected in 1) with those of a \textit{Spitzer} sample of O and B-type stars \citep{Gordon2021}} following the method of~\citet{Gordon2021}. Stars occupy different loci of the $J-K_S$ vs. $K_S - \mathrm{MIPS24}$ color-color diagram~\citep[Fig. 4. in ][]{Gordon2021} according to the amount of reddening along their lines of sight and the presence of stellar winds. Fig. \ref{fig:sourceselection} shows the $J-K_S$ vs. $K_S - W_4$ color-color diagram of the sources selected in (1). The $W4$ band is similar to the MIPS24 band. 
     We applied the following selection criteria to select reddened stars and avoid stars with strong winds and/or emission lines as much as possible: 
     \begin{equation}%\label{eq:sel_criteria}
     \begin{split}
     K_S - W_4 < 0.6 \, \mathrm{mag};  \hspace{0.1cm} \\
      J - K_S > 0.6 \, \mathrm{mag}; \hspace{0.1cm} \\
      H - K_S > 0.2 \, \mathrm{mag}.
     \end{split}
      \label{eq:one}
     \end{equation}

     \item{We required our sample to contain stars with line-of-sight visual extinctions $A_\mathrm{V} > 4.0$
mag (Section~\ref{sec:feasibility}): Such stars should have infrared dust absorption features (both intrinsically strong and weak) that are strong enough to be detected and characterized (see section~\ref{sec:feasibility}).} % more info about this }
\end{enumerate}

\begin{figure*}[ht!]
\centering
\includegraphics[width=\textwidth]{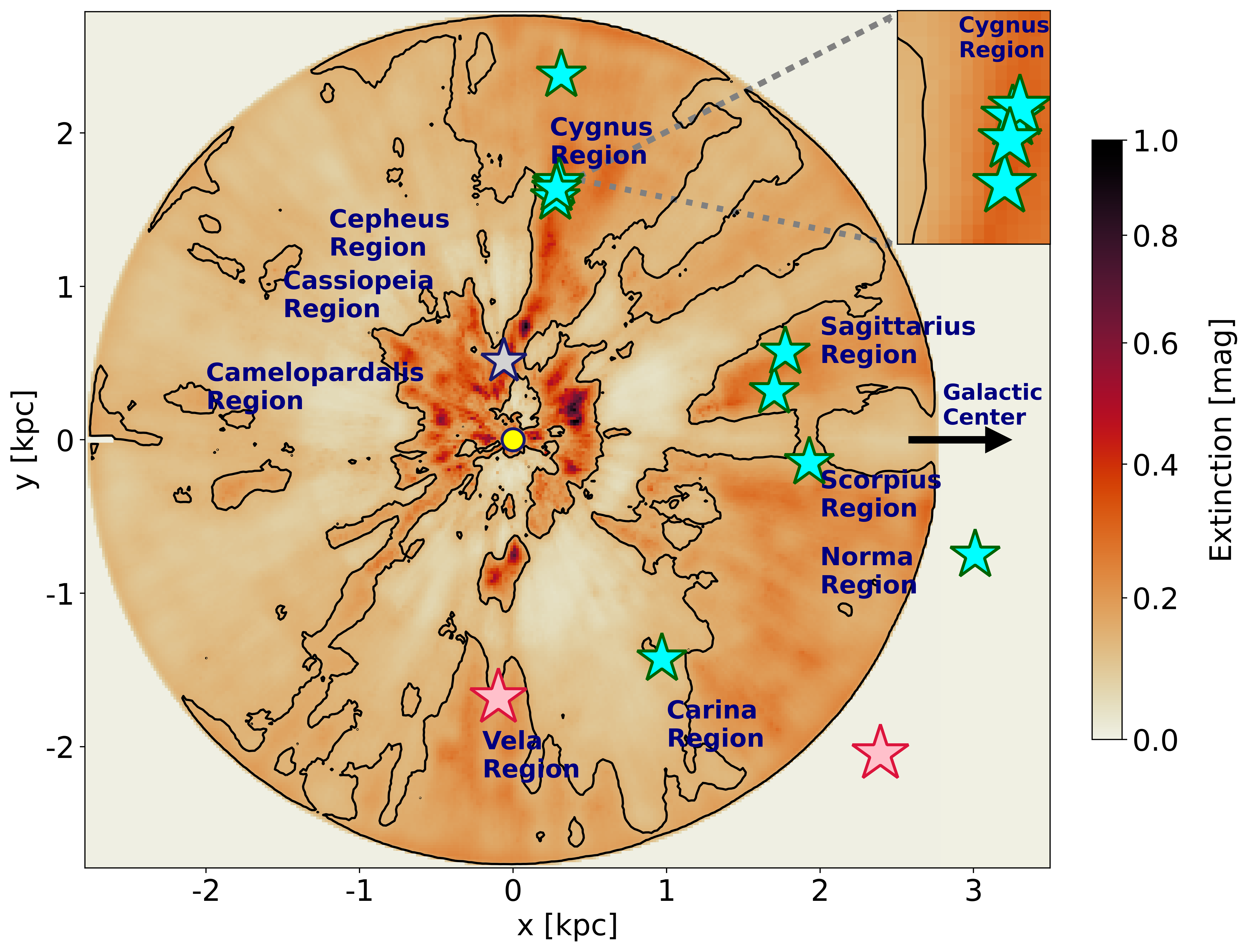}
\includegraphics[width=\textwidth]{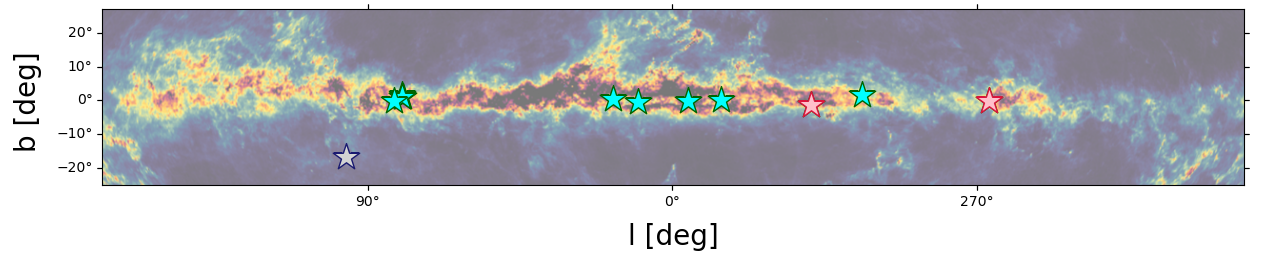}
\caption{\textit{Top}: Extinction map of the local ISM viewed top-down taken from \citet{Dharmawardena24}. The Sun (yellow circle) is at $X$, $Y$ = 0, 0 and the Galactic centre is at $X$, $Y$ = 8.2, 0 kpc. The twelve targets (cf. Table \ref{tab:sources}) are represented by colored stars. The two stars analyzed in this paper are shown in red and the other ten are shown in blue. The small gray star near the center of this figure indicates 10 Lac ($A_{\rm V} = 0.21~$mag), which was used as a reference star. The contours represent extinction in steps of 0.2 mag. 
\textit{Bottom}: Dust map of the Milky Way galactic plane derived from Gaia photometry (see \citet{Andrae:23} for more details) with the twelve targets following the same color scheme as above. Note that some of the targets overlap in this projection.
\label{fig:map}}
\end{figure*}

These criteria led to our selection of 11 sources whose sight lines are grouped in three regions: toward Cygnus, Sagittarius-Scorpius, and Norma-Vela, as shown in Figure \ref{fig:map}. The sources in the Sagittarius-Scorpius region are broadly located in the direction of the Galactic center.
We additionally included VI Cyg 1 which, having being observed by \textit{Spitzer} IRS \citep{Misselt2008}, allows for a direct comparison of spectral features across instruments and decades. Table~\ref{tab:sources} provides an overview of these 12 targets, containing their coordinates, distances, spectral types, V band magnitudes, $\mathrm{B}-\mathrm{V}_0$ and E(B-V). 
In Figure~\ref{fig:map} we also indicate the position of the star 10 Lacertae (10 Lac). This source is used to compare our sources with an unobscured sightline, since the extinction along the line of sight toward this calibration star~\citep{Law2025} is very low ($A_{\rm V} = 0.21~$mag).

\begin{table*}
\caption{Properties of the WISCI targets, this and previous studies. \label{tab:sources}}
\centering
\def\arraystretch{1.1}
\begin{tabular}{l | l | c c l c c c c c}
 \hline
2MASS J[...] & Name in proposal & RA  Dec & $l$ $b$ & Sp. Type & V & (B-V)$_{0}$ & E(B-V) & Ref. & Distance \\
   & id 2183 & [deg] & [deg] &  & [mag] & [mag] & [mag] & & [kpc]  \\
 \hline\hline
08574757$-$4609145 & GSC 08152-02121 &  134.44  $-$46.15 & 266.77 $-$0.34 & B2~IV & 13.14 & -0.21 & 1.66 & 1 & $1.68^{+0.09}_{-0.08}$ \\
13015278$-$6131045 & TYC 8989-436-1 & 195.46 $-$61.52 & 304.18 +1.33 & B0.5~II & 11.28 & -0.20 & 1.71 & 1 & $1.73^{+0.40}_{-0.40}$ \\
15095841$-$5958463 & CPD-59 5831  & 227.49 $-$59.98 & 319.46 $-$1.63 & B5~Ia & 10.27 & -0.08 & 1.31 & 1 & $3.15^{+0.11}_{-0.14}$  \\
17075654$-$4040383 & CD-40 11169 & 256.99 $-$40.68 & 345.96 $-$0.15 & B8~Ia & 10.88 & -0.03 & 1.61 & 1 & $3.10^{+0.18}_{-0.20}$ \\
17362876$-$3253166 & TYC 7380-1046-1 & 264.12 $-$32.89 & 355.61 $-$0.43 & B0~Ia & 10.70 & -0.22 & 1.40 & 1 & $1.94^{+0.08}_{-0.07}$  \\ 
18112983$-$2017075 & TYC 6272-339-1 & 272.87 $-$20.29 & 10.43 $-$0.76 & B3~II & 11.24 & -0.13 & 0.93 & 1 & $1.73^{+0.06}_{-0.04}$ \\
18230252$-$1320387 & LS 4992  & 275.76 $-$13.34 & 17.85 +0.10 & O8~Iaf & 11.26 & -0.30 & 1.69 & 1& $1.86^{+0.17}_{-0.15}$ \\
20311055$+$4131535 & VI Cyg 1 & 307.79 $+$41.53 & 80.17 +1.23 & O8~V +B & 11.09 & -0.285 & 1.89 & 2 & $1.69^{+0.05}_{-0.06}$ \\
20323486$+$4056174 & [CPR2002] A38  & 308.14 $+$40.94 & 79.85 +0.67 & O8~V & 12.35 & -0.285 & 2.87 & 3 &$1.70^{+0.04}_{-0.04}$ \\
20331106$+$4110321 & ALS 15181 & 308.30 $+$41.17 & 80.11 +0.72 & B0~V & 14.78 & -0.26 & 2.45 & 4 & $1.60^{+0.03}_{-0.03}$ \\ 
20332674$+$4110595 & GSC 03157-00327 & 308.36 $+$41.18 & 80.14 +0.68 & O8~Vz & 13.00 & -0.285 & 2.16 & 5 &  $1.66^{+0.05}_{-0.03}$  \\ 
20452110$+$4223513 &  & 311.34 $+$42.40 & 82.46 $-$0.34 & B2~V & 12.80 & -0.21 & 2.69 & 6 & $2.40^{+0.09}_{-0.10}$ \\
 \hline
\end{tabular}
\raggedright
\tablecomments{[CBJ]: distances taken from \citet{Bailer21}, RA and Dec: ICRS coordinates (epoch J2000)}
\tablerefs{1. This study; 2. \cite{2008ApJ...679.1478K}; 3. \cite{2008A&A...487..575N}; 4. \citet{2007ApJ...664.1102K}; 5. \cite{2016ApJS..224....4M}; 6. \cite{2012A&A...543A.101C}.}
\end{table*}

\subsection{Overview of observations}
\label{ssec:obs}
 
This section provides an overview of all observations of the twelve stars used in the WISCI program, which includes photometry and spectral observations from ultraviolet to mid-infrared wavelengths.
The core of the WISCI project are the \textit{JWST} NIRCam and MIRI Medium Resolution Spectroscopy (MRS) spectra, covering near- and mid-infrared wavelengths wherein the majority of the characteristic features for diffuse ISM dust can be found. In addition to these, stellar spectral typing was assisted by optical spectra from VLT XSHOOTER (for southern targets). 
We further took photometric measurements from the literature to serve as a point of comparison with these newly acquired spectroscopic data sets.
The brightest, and least extinguished, stars in the sample also have \textit{HST} STIS ultraviolet observations revealing the bump at 2175~\AA{}\ for comparison with the strength of carbonaceous dust features at near-infrared wavelengths. The continuous coverage of VLT XSHOOTER, HCT HFOSC and HESP, and \textit{HST} STIS across optical wavelengths enables the search for diffuse interstellar bands (DIBs), many of which are shallow at the $A_{\rm V}$ levels probed by WISCI. 

A detailed overview of the observations and data reduction of all targets in the WISCI program is presented in Appendix \ref{app:obs_summary} including the program IDs for all observations \ref{tab:obs_summary}. The data sets available for each WISCI target are heterogeneous; however, the processing of the data was consistent for every target observed by a given facility/instrument. 

\section{Analysis and Results} 
\label{sec:res}

\subsection{Spectral classification of the WISCI sample}
\label{sec:star_properties}

The spectral types of seven of the WISCI targets were not known with great precision prior to their inclusion in the sample, being predominantly based on photometry. Whilst sufficient for a determination of feasibility for the project, their classifications were inadequate for the fine detail work necessary to extract subtle features imparted by dust from the observations. The remaining five had been classified spectroscopically in previous studies.  

\begin{figure*}[ht!]
    \includegraphics[width=\textwidth]{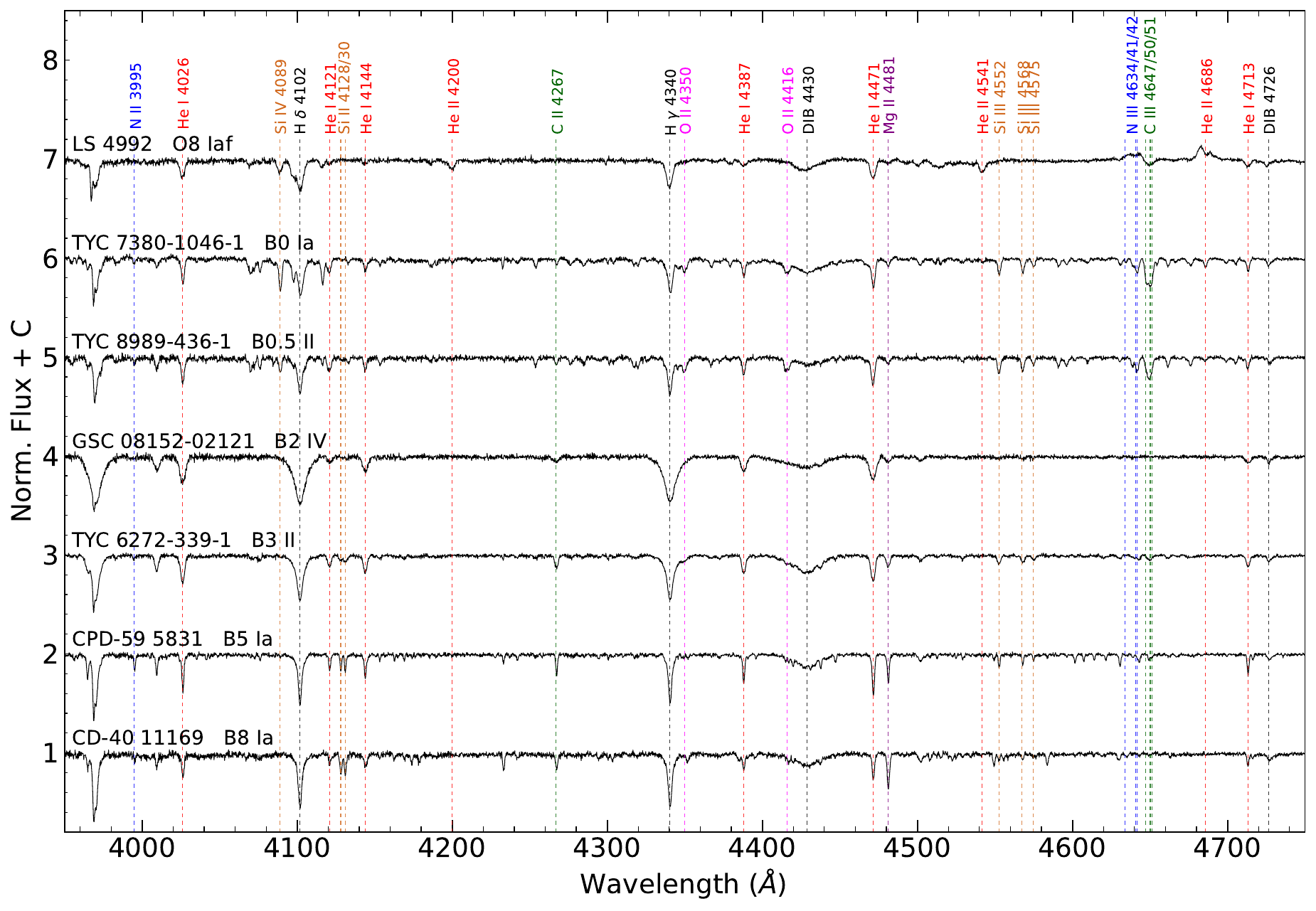}
    \caption{VLT XSHOOTER optical spectra of seven targets spanning 3950 to 4750 \AA{}. The target spectra have been normalized to their respective continua and vertically offset from each other for clarity. The solid black line denotes the observed spectrum of each star. Vertical dashed lines and labels denote various lines important to diagnose the stellar spectral types. In addition, we highlight a strong diffuse interstellar band at 4430 \AA{}, which is present in all the spectra.
    \label{fig:spectypevlt}}
\end{figure*}

We therefore pursued a program of characterization for the WISCI sample to refine the target spectral types. To this end we acquired and reduced VLT XSHOOTER observations of the targets, as described in Appendix~\ref{app:obs_summary}. 
We classified the WISCI stars based on the strength of their H, He, and metal lines in the UVB spectrum for the seven targets with VLT observations (Fig. \ref{fig:spectypevlt}). We refer to the criteria from \citet{2019ApJS..241...32L}, \citet{2009ssc..book.....G}, and \citet{2011ApJS..193...24S}.

\begin{itemize}
  \item \emph{LS 4992 (O8 Iaf)}: This is an O star spectrum. The He II lines at 4200, 4541 and 4686 \AA{} are strong. We note that He I 4471 $>$ He II 4541, which means that this is an O8 to O9 star according to \citet{2019ApJS..241...32L}. He II 4541 $>$ He I 4387 and He II 4200 $>$ He I 4144 further prove this is a O8 star according to \citet{2011ApJS..193...24S}. He II 4686 is strongly in emission, which indicates a spectral type of O8 Ia. We also observe strong N III 4641-42 emission further suggesting that LS 4992 is an O8 Iaf star.
  
  \item \emph{TYC 7380-1046-1 (B0 Ia)}: Weak He II lines are present in the spectrum, indicating an early B type star. To distinguish the subtype from B0 to B1, we can use the ratio Si IV 4089/Si III 4552 according to \citet{2009ssc..book.....G}. The ratio is about two in this spectrum, indicating a subtype of B0. The O II lines are quite strong in early B type stars and their strengths increase with luminosity. We use their ratio with He lines to determine the luminosity class. He I 4387 and O II 4416 almost have the same strengths, which leads to a luminosity class of supergiant. The ratio Si IV 4552/He I 4387 further results in a spectral type of B0 Ia.
  
  \item \emph{TYC 8989-436-1 (B0.5 II)}: The He II lines are weaker compared to the B0 star above. The ratio Si IV 4089/Si III 4552 is close to unity, which indicates a B0.5 subtype. Si III 4552 $\approx$ Si III 4568 and Si III 4568 $>$ Si III 4575, resulting in a luminosity class between supergiant and giant. O II 4350 with intermediate strength further confirms a spectral type of B0.5 II.
  
  \item \emph{GSC 08152-02121 (B2 IV)}: He II lines are absent from the spectrum, therefore the spectral class is later than B0.7. Si II 4128-4130 is much weaker than He I 4121 and when considering the ratio He I 4471/Mg II 4481, the spectral type is B2. The broad H and He lines together with the weak O II and N II lines indicate a luminosity class of subgiant to dwarf. The ratio He I 4121/H $\delta$ finally results in a spectral type of B2 IV.
  
  \item \emph{TYC 6272-339-1 (B3 II)}: The Mg II 4481 and Si II 4128-30 lines are stronger than in the B2 stars and demonstrate that the spectral type of TYC 6272-339-1 is B3. The strength of N II 3995 indicates a luminosity class between bright giant and giant. The ratios He I 4121/H$\delta$ and He I 4387/H$\gamma$ point towards B3 II.
  
  \item \emph{CPD-59 5831 (B5 Ia)}: Si II 4128-4130 is stronger than He I 4121, while Mg II 4481 remains weaker than He I 4471, indicating a B5 spectral type. The ratio Si III 4522/He I 4387 and the strong N II 3995 line are consistent with B5 Ia.
  
  \item \emph{CD-40 11169 (B8 Ia)}: Mg II 4481 is stronger than He I 4471 and their ratio results in a subtype of B8. The H and He lines are quite narrow, indicating a low surface gravity. We notice that Si II 4128-30 is stronger than He I 4144. From consideration of the strengths of the He I lines, we conclude that this is a B8 Ia star.
\end{itemize}

The remaining five stars can only be observed from the northern hemisphere and thus do not have VLT spectra. The spectral resolution of the \textit{HST} does not allow detailed spectral classification. However, analysis of the spectral types of these remaining five stars were available before the start of this project and we refer to these studies. 

\begin{itemize}
  \item \emph{GSC 03157-00327 (O8 Vz)}: GSC 03157-00327 is indicated as ALS 15134 in \citet{2016ApJS..224....4M}; they classified it as spectral type O8 Vz. 
  
  \item \emph{$[$CPR2002$]$ A38 (O8 V)}: \citet{2008A&A...487..575N} presented a spectrum of it and classified it as an O8 V star. 
  
  \item \emph{VI Cyg 1 (O8 V + B)}: \citet{2008ApJ...679.1478K} conclude that VI Cyg 1 is a single-lined spectroscopic binary based on analysis of the multi-epoch spectra and radial velocities. The primary is an O8 V star and the secondary is a B type star. VI Cyg 1 is indicated as MT 059 in~\citet{2008ApJ...679.1478K}
  
  \item \emph{ALS 15181 (B0 V)}: ALS 15181 is indicated as [MT91] 435 in \citet{2007ApJ...664.1102K} where they classified it as a B0 V star. 
  
  \item \emph{2MASS J20452110+4223513 (B2 V)}: \citet{2012A&A...543A.101C} presented a spectrum of it and classified it as a B2 V star. 
\end{itemize}

\subsection{Extinction}
\label{sec:extinction}

The line-of-sight extinction properties were measured using the Python package {\sc measure\_extinction}\footnote{\url{https://github.com/karllark/measure_extinction}} \citep{karl_gordon_2024_11186967} and are presented in Table~\ref{tab:extinction}.
The extinction curves were calculated by comparison of each reddened target star to the appropriate unreddened stellar atmosphere model \citep[e.g.,][]{Fitzpatrick19}.
The appropriate stellar atmosphere model was determined by modeling the \textit{HST} STIS optical spectrum, and optical $BV$ and near-infrared $JHK_{S}$ photometry \citep{2014HendenMunari,2006Skrutskie} using Tlusty OB stellar atmosphere models (Hubeny et al. submitted) extinguished with an $R(V)$ dependent extinction model appropriate for the Milky Way \citep{Gordon09FUSE, Fitzpatrick19, Gordon2021, Decleir22, 2023Gordon}.
We then calculate the maximum likelihood interstellar extinction model for each target using the Markov Chain Monte Carlo (MCMC) code \textit{emcee} \citep{2013ForemanMackey}. The stellar spectral types from \cite{2024Wei} were used as priors for the fitting with Gaussian priors on the stellar parameters $T_\mathrm{eff}$ ($\sigma = 2000$~K) and $\log(g)$ ($\sigma = 0.1$).
The translation of spectral types to $T_\mathrm{eff}$ and $\log(g)$ is different for each stellar atmosphere grid, hence the use of the spectral types as a prior.
The fitting included a nuisance parameter ($\ln f$) to account for underestimating the observational uncertainties.
The log-probability of each realization of the model is calculated using the error-weighted least-squares fit of the stellar model to the observations, using the Python package {\sc Synphot}\footnote{\url{https://synphot.readthedocs.io/en/latest/}} \citep{Synphot} to calculate synthetic photometry from the model spectra. 
We explore the parameter space using 50 walkers and 5\,000 steps per target (250\,000 model realizations). The posterior probability distribution is then constructed from the final 50,000 realizations of the model. The parameter values for $T_{\rm eff}$, $\log g$, $A_{\rm V}$, and $R_{\rm V}$ and their uncertainties are then determined from the 16th, 50th, and 84th percentiles of the marginalized posterior probability distributions.

Whilst we determine values for $A(V)$ and $R(V)$ (in addition to $T_\mathrm{eff}$ and $\log(g)$) from that process, these values are tied to the assumption of the \citep{Gordon2021} extinction model.
The extinction properties were determined using the standard method of extrapolating the un-normalized extinction curve to infinite wavelength \citep{Martin90, Gordon2021, Decleir22}.
The optical and NIR extinction curve ($E(\lambda - 55)$, where 55 indicates 0.55 \micron) was calculated and extrapolated assuming the \citet{Rieke85} optical/IR extinction curve shape to determine the dust column $A(55)$.
The extinction curve was referenced to the spectroscopic observations at 0.55~\micron\ as this provides a measurement unaffected by bandpass effects \citep{Fitzpatrick19}.
Combining $A(55)$ with the measured color excess $E(44-55)$, the ratio of total-to-selective extinction $R(55)$ was calculated.
The values of $R(V)$ are very similar and can be calculated using $R(V) = 1.01\,(R(55) + 0.049)$ \citep[Eq.~5 of][]{Fitzpatrick19}.
The $A(V)$ value can be computed from $A(V) = A(55) - 0.049\,E(44-55)$ that can be derived from Eq.~4 of \citet{Fitzpatrick19}.
The values of $R(V)$ are all quite similar and near the Milky Way average of $\sim$3.1.
This is not surprising as these are high $A(V)$ sight lines that likely average over variations in $R(V)$.
It is reassuring that the sight lines that probe dust in a similar region (e.g., those 2MASS~J203xx sight lines) show similar $R(V)$ values (e.g., $\sim$2.9).

All of the fitting and calculations are provided through the {\sc empirical\_extinction} 
repository\footnote{\url{https://github.com/WISCI/empirical_extinction}} on the WISCI GitHub.
The detailed study of the NIR--MIR continuum and feature extinction and relation to the same in the optical and UV extinction will be presented in future work.

\begin{table*}
    \caption{Extinction Properties for the WISCI sample. The two targets presented in this paper are highlighted in bold. \label{tab:extinction}}
    \centering
    \begin{tabular}{lccccc}
     \hline
     2MASS J[...]  & $E(55-44)$ & $A(55)$ & $R(55)$ & $A(V)$\tablenotemark{a} & $R(V)$\tablenotemark{a} \\
     \hline\hline
     \textbf{08574757$-$4609145} & $1.60\pm0.02$ & $4.84\pm0.02$ & $3.03\pm0.04$ & $4.76 \pm 0.02$ & $3.11 \pm 0.04$ \\ 
     13015278$-$6131045 & $1.81\pm0.01$ & $5.62\pm0.02$ & $3.10\pm0.02$ & $5.53 \pm 0.02$ & $3.18 \pm 0.02$ \\
     \textbf{15095841$-$5958463} & $1.53\pm0.01$ & $4.56\pm0.02$ & $2.99\pm0.02$ & $4.49 \pm 0.02$ & $3.07 \pm 0.02$ \\
     17075654$-$4040383 & $1.81\pm0.01$ & $5.38\pm0.02$ & $2.97\pm0.02$ & $5.29 \pm 0.02$ & $3.05 \pm 0.02$ \\ 
     17362876$-$3253166 & $1.57\pm0.05$ & $5.07\pm0.03$ & $3.23\pm0.11$ & $4.99 \pm 0.03$ & $3.31 \pm 0.11$ \\  
     18112983$-$2017075 & $1.33\pm0.05$ & $4.52\pm0.02$ & $3.40\pm0.13$ & $4.45 \pm 0.02$ & $3.48 \pm 0.13$ \\ 
     18230252$-$1320387 & $1.77\pm0.05$ & $5.59\pm0.02$ & $3.16\pm0.09$ & $5.50 \pm 0.02$ & $3.24 \pm 0.09$ \\ 
     20311055$+$4131535 & $1.70\pm0.05$ & $5.05\pm0.02$ & $2.97\pm0.09$ & $4.97 \pm 0.02$ & $3.05 \pm 0.09$ \\ 
     20323486$+$4056174 & $2.30\pm0.07$ & $6.35\pm0.02$ & $2.76\pm0.09$ & $6.24 \pm 0.02$ & $2.84 \pm 0.09$ \\
     20331106$+$4110321 & $2.41\pm0.11$ & $6.91\pm0.03$ & $2.86\pm0.13$ & $6.79 \pm 0.03$ & $2.94 \pm 0.13$ \\  
     20332674$+$4110595 & $2.20\pm0.06$ & $6.29\pm0.02$ & $2.86\pm0.08$ & $6.18 \pm 0.02$ & $2.94 \pm 0.08$ \\  
     20452110$+$4223513 & $2.16\pm0.08$ & $6.28\pm0.03$ & $2.91\pm0.10$ & $6.17 \pm 0.03$ & $2.99 \pm 0.10$ \\ 
    \hline
    \end{tabular}
\tablenotetext{a}{Calculated from $E(44-55)$, $A(55)$, and $R(55)$ as described in the text.}
\end{table*}

\subsection{Absorption features in the spectra of GSC 08152-02121 and  CPD-59 5831}
\label{sec:dust_properties}

\begin{figure*}[ht!]
    \includegraphics[width=\textwidth]{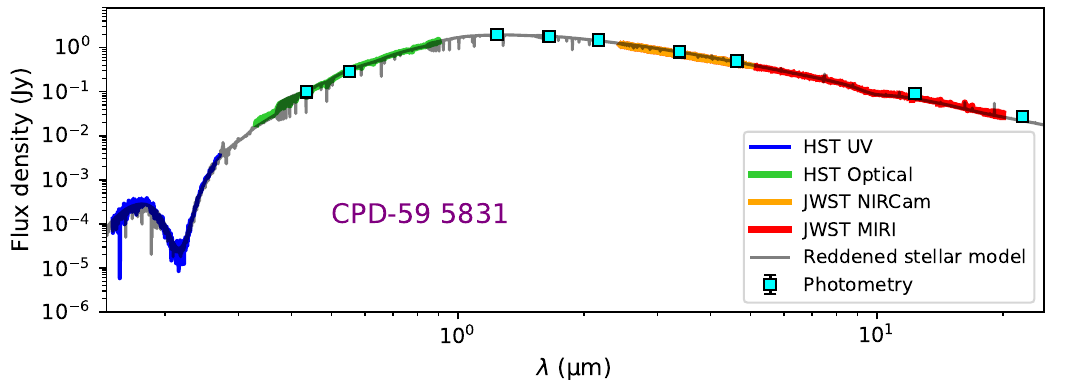}
    \includegraphics[width=\textwidth]{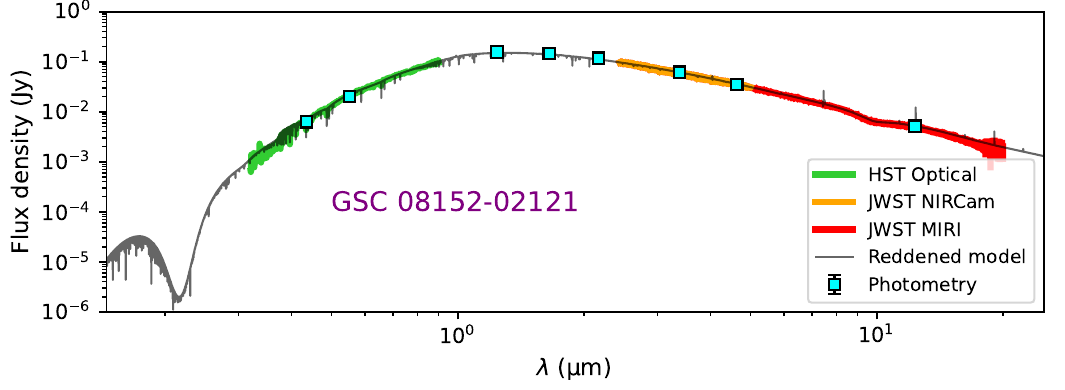}
    \caption{Spectral energy distributions of CPD-59 5831 (top; 2MASS J15095841-5958463) and GSC 08152-02121 (bottom; 2MASS J08574757-4609145). The reddened Tlusty theoretical stellar photosphere models are shown as gray solid lines. The \textit{HST} and \textit{JWST} spectra are denoted by blue (UV), green (optical), orange (NIRCam), and red (MRS) lines, respectively. Photometry from APASS ($BV$), 2MASS ($JHK_{\rm S}$) and \textit{WISE} ($W1$-$W4$) are denoted by blue squares.
    \label{fig:sed_extinction}}
\end{figure*}

The observed spectra of GSC 08152-02121 (2MASS 08574757-4609145) and CPD-59 5831 (2MASS J15095841-5958463) are shown in Figure~\ref{fig:sed_extinction}, where we present the \textit{HST} STIS and \textit{JWST} NIRCam and MIRI data. The spectrum of CPD-59 5831 is $\sim10$ times brighter than GSC 08152-02121 and allows for observation of the 2175~\AA\ feature. In both spectra we can observe diffuse interstellar bands, e.g. the 4430~\AA\ DIB in Figure~\ref{fig:spectypevlt}. 
In this section we focus on the modeling of their infrared ISM dust absorption features
 using the Python-based {\sc Astropy} modeling routines, {\sc Scipy} fitting routines \citep{astropy2013,astropy2018,astropy2022,SciPy} and \texttt{emcee} \citep{2013ForemanMackey}. The code developed to analyze these dust features is available on Github\footnote{\url{https://gitfront.io/r/SZeegers/FjeYBBQsUL7n/WISCIpaper1/}}. The feature model consists of two components, combining a power law to represent the local continuum around each feature and a set of one or more (skewed) Gaussian components to represent the dust absorption. 
Stellar H and He spectral lines within the regions of interest for each ISM feature were identified using the NIST database\footnote{\url{https://www.nist.gov/pml/atomic-spectra-database}} \citep{NIST_DB}.

\subsubsection{Carbonaceous dust features}
\label{sec:carbon_irfeatures}
% upper limits on nano diamonds?
 
In order to investigate the aliphatic and aromatic hydrocarbons, we fit the absorption features around 3.4, 6.2, 6.8 and 7.2~\micron. We fit a local continuum using a power law to minimize the effect of absorption unrelated to the hydrocarbons. Since these features are weak, we use a range of local continua to mitigate possible biases a single continuum selection may induce. The continuum is varied from the largest possible wavelength range to the smallest.
We further account for the presence of stellar lines within the wavelength range of interest by using Gaussian models with fixed peak wavelengths (taken from the NIST database) and arbitrary widths (standard deviations between 0.001 and 0.01\micron, as free parameters) to represent (and subtract) their contribution to the sought-after dust absorption features.

\begin{table*}[]
    \centering
    \caption{Summary of carbonaceous dust feature wavelengths and depths identified in the \textit{JWST} spectra. Wavelength denotes the deepest point of the feature, the depth is relative to a fitted continuum at the given wavelength. Fixed values for the feature wavelengths and FWHM are taken from~\citet{Chiar13}. \label{tab:carbonir_features}}
    \renewcommand{\arraystretch}{1.4}
    \begin{tabular}{llcccc}
        \hline
       Source & Feature & Wavelength & Optical depth  & FWHM & Integrated Area \\
                &    & (\micron) &       & (\micron) & ($\mathrm{cm}^{-1}$)   \\
        \hline
        GSC & $\mathrm{CH}$ aromatic &  3.289 & 
        $0.003^{\pm0.001\,\mathrm{(stat)}}_{\pm0.004\,\mathrm{(sys)}}$ & $0.090$ (fixed)&  $0.310^{\pm 0.110\,\mathrm{(stat)}}_{\pm 0.310\,\mathrm{(sys)}}$ \\       
          
       08152-02121  & $\mathrm{CH}_3$ aliphatic & 3.376 & $0.008^{\pm0.001\mathrm{(stat)}}_{\pm0.002\,\mathrm{(sys)}}$ & $0.050$ (fixed)& $0.366^{\pm 0.068\,\mathrm{(stat)}}_{\pm 0.080\,\mathrm{(sys)}}$ \\

   &$\mathrm{CH}_2$ aliphatic & $3.420$ & $0.012^{\pm0.001\,\mathrm{(stat)}}_{\pm0.002\,\mathrm{(sys)}} $ &  $0.050$ (fixed)& $0.562^{\pm 0.068\,\mathrm{(stat)}}_{\pm 0.087\,\mathrm{(sys)}}$\\

        & $\mathrm{CH}_3$ aliphatic & 3.474 & $0.013^{\pm0.001\,\mathrm{(stat)}}_{\pm0.002\,\mathrm{(sys)}}$ & $0.050$ (fixed)& $0.586^{\pm 0.065\,\mathrm{(stat)}}_{\pm 0.077\,\mathrm{(sys)}}$\\
        & $\mathrm{CH}_2$ aliphatic & 3.520 & $0.009^{\pm0.001\,\mathrm{(stat)}}_{\pm0.002\,\mathrm{(sys)}}$ &$0.050$ (fixed)& $0.380^{\pm 0.062\,\mathrm{(stat)}}_{\pm 0.093\,\mathrm{(sys)}}$ \\
        & CC olefinic & 6.19 & $0.011^{\pm0.002\,\mathrm{(stat)}}_{\pm0.001\,\mathrm{(sys)}}$ & $0.061^{\pm{0.009\,\mathrm{(stat)}}}_{\pm0.003\,\mathrm{(sys)}}$ & $0.182^{\pm0.037\,\mathrm{(stat)}}_{\pm0.022\,\mathrm{(sys)}}$  \\
        & CC aromatic & 6.25 & $0.003^{\pm0.002\,\mathrm{(stat)}}_{\pm0.001\,\mathrm{(sys)}}$ & $0.037^{\pm0.008\,\mathrm{(stat)}}_{\pm0.001\,\mathrm{(sys)}}$ & $0.131^{\pm0.067\,\mathrm{(stat)}}_{\pm0.082\,\mathrm{(sys)}}$ \\
        & $\mathrm{CH}_2$ aliphatic & 6.85 & $<0.005$ & $<0.25$ & $<0.26$  \\ 
        & $\mathrm{CH}_3$ aliphatic & 7.25 & $<0.002$  & $<0.19$ & $<0.08$    \\    
        
        \hline
        CPD-59 5831 & $\mathrm{CH}$ aromatic & 3.289 & $0.008^{\pm0.001\,\mathrm{(stat)}}_{\pm0.001\,\mathrm{(sys)}}$ & $0.090$ (fixed)& $0.699^{\pm 0.126\,\mathrm{(stat)}}_{\pm 0.093\,\mathrm{(sys)}}$\\ 
       & $\mathrm{CH}_3$ aliphatic & 3.376 & $0.006^{\pm0.002\,\mathrm{(stat)}}_{\pm0.001\,\mathrm{(sys)}}$ & $0.050$ (fixed) & $0.292^{\pm 0.085\,\mathrm{(stat)}}_{\pm 0.027\,\mathrm{(sys)}}$\\
        & $\mathrm{CH}_2$ aliphatic & 3.420 & $0.009^{\pm0.002\,\mathrm{(stat)}}_{\pm0.001\mathrm{(sys)}}$ & $0.050$ (fixed) &  $0.430^{\pm 0.083\,\mathrm{(stat)}}_{\pm 0.028\,\mathrm{(sys)}}$ \\
        & $\mathrm{CH}_3$ aliphatic & 3.474 & $0.006^{\pm0.002\,\mathrm{(stat)}}_{\pm0.001\,\mathrm{(sys)}}$ & $0.050$ (fixed) & $0.260^{\pm 0.080\,\mathrm{(stat)}}_{\pm 0.024\,\mathrm{(sys)}}$  \\
        &$\mathrm{CH}_2$ aliphatic & 3.520 & $0.004^{\pm0.002\,\mathrm{(stat)}}_{\pm0.001\,\mathrm{(sys)}}$ & $0.050$ (fixed)& $0.171^{\pm 0.076\,\mathrm{(stat)}}_{\pm 0.028\,\mathrm{(sys)}}$\\
        & CC olefinic & 6.19 & $0.010^{\pm0.001\,\mathrm{(stat)}}_{\pm0.001\,\mathrm{(sys)}}$ & $0.091^{\pm0.015\,\mathrm{(stat)}}_{\pm0.012\,\mathrm{(sys)}}$ & $0.257^{\pm0.056\,\mathrm{(stat)}}_{\pm0.049\,\mathrm{(sys)}}$  \\
        & CC aromatic & 6.25 & $0.005^{\pm0.002\,\mathrm{(stat)}}_{\pm0.001\,\mathrm{(sys)}}$ & $0.014^{\pm0.006\,\mathrm{(stat)}}_{\pm0.001\,\mathrm{(sys)}}$ & $0.274^{\pm0.079\,\mathrm{(stat)}}_{\pm0.062\,\mathrm{(sys)}}$\\
        & $\mathrm{CH}_2$ aliphatic & 6.85 & $<0.001$ & $<0.24$  & $<0.04$  \\ 
        & $\mathrm{CH}_3$ aliphatic & 7.25 & $<0.002$ &  $<0.15$ & $<0.04$ \\          

         \hline
    \end{tabular}
\end{table*}

\begin{figure*}
    \centering
    \includegraphics[width=0.48\textwidth]{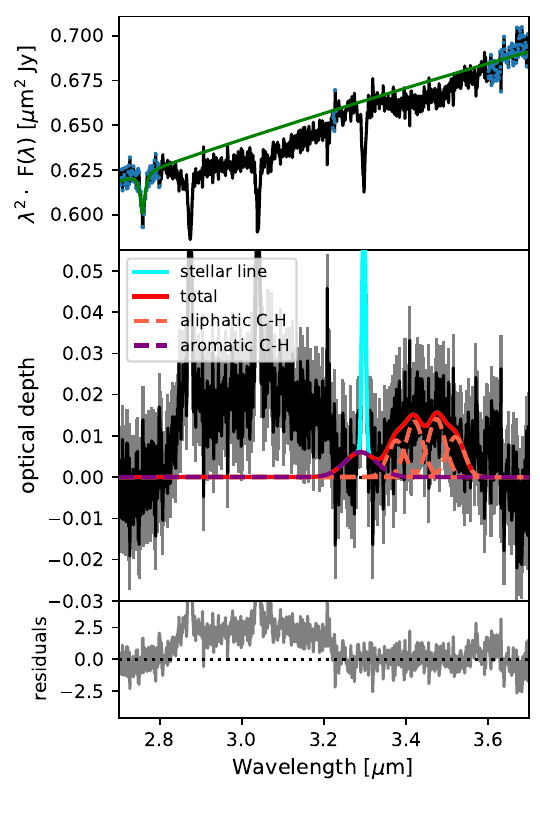}
    \includegraphics[width=0.48\textwidth]{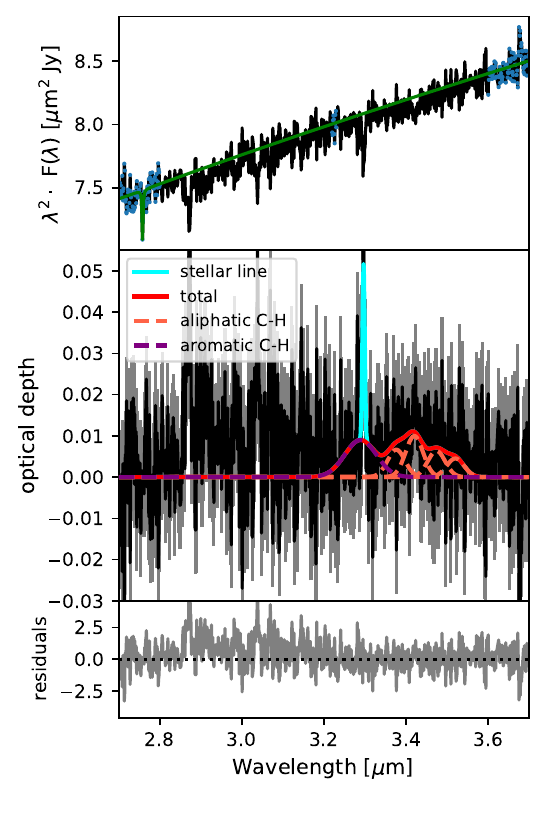}
    \caption{NIRCam spectra of the 3.3~\micron\ aromatic and 3.4~\micron\ aliphatic hydrocarbon features for GSC 08152-02121 (left) and CPD-59 5831 (right). In the top panel we show the observed spectrum in black, blue dots denote data points from which the continuum was calculated, and the fitted continuum as a green line. The middle panel shows the Gaussian components of the dust model as orange (aliphatic) and purple (aromatic) dashed lines, with the total denoted by a red solid line, all in units of optical depth. The stellar line (Hydrogen Pfund-$\delta$) contaminating the model fit at 3.3 $\micron$ is denoted by a cyan solid line. The gray contours behind the data indicate the $1\sigma$ statistical uncertainty derived from the continuum. The bottom panel shows the significance of the residuals ($\pm2.5\sigma$) after subtraction of the continuum and line model, and normalization by the standard deviation.
    Shortward of the 3.3 $\micron$ aromatic and 3.4 $\micron$ aliphatic hydrocarbon features the GSC 08152-02121 spectrum shows an absorption feature which may be attributed to trapped water. No comparable absorption is seen at the same wavelength range in the CPD-59 5831 spectrum.}
    \label{fig:c34_absorption}
\end{figure*}

\begin{figure*}
    \centering
    \includegraphics[width=0.48\textwidth]{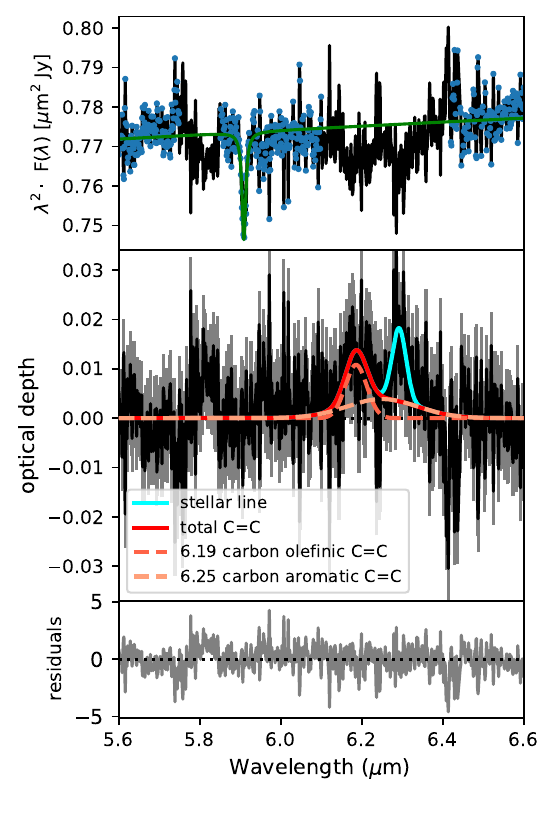}
    \includegraphics[width=0.48\textwidth]{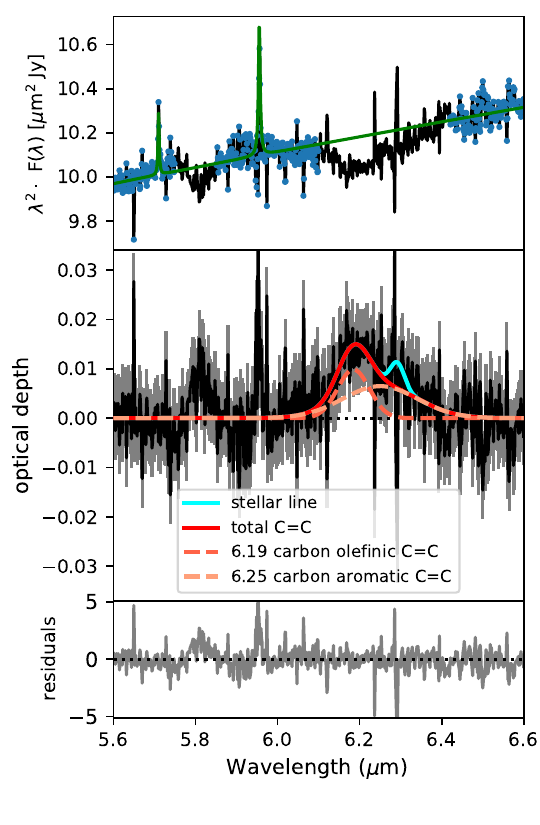}
    \caption{Here we show the MIRI MRS spectrum around 6.2~\micron\ for GSC 08152-02121 (left) and CPD-59 5831 (right). In the top panel we show the observed spectrum in black with the fitted continuum as a green line. Blue data points denote the parts of the spectrum used to determine the continuum. An unidentified feature is present around 5.8~\micron\ which is excluded from the continuum. The middle panel shows the Gaussian dust components of the model fit in units of optical depth. The two components are shown by dashed lines and the total dust component is shown by the red line. The cyan line indicates the position and strength of a stellar line (Hydrogen $n=7$ to $n=13$). The gray contours behind the data indicate the $1\sigma$ statistical uncertainty derived from the continuum. The bottom panel shows the significance of the residuals ($\pm5\sigma$) after subtraction of the continuum and line model, and normalization by the standard deviation.}
    \label{fig:c62_absorption}
\end{figure*}

\begin{figure*}
    \centering
    \includegraphics[width=0.48\textwidth]{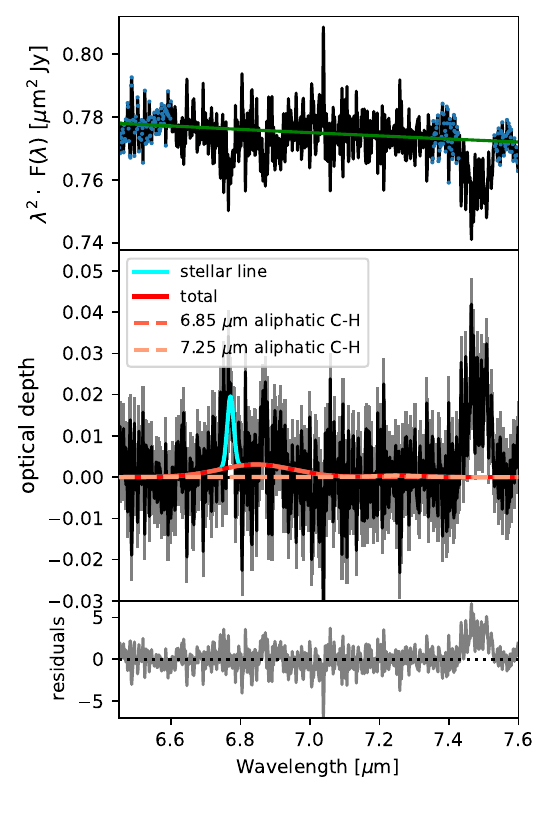}
    \includegraphics[width=0.48\textwidth]{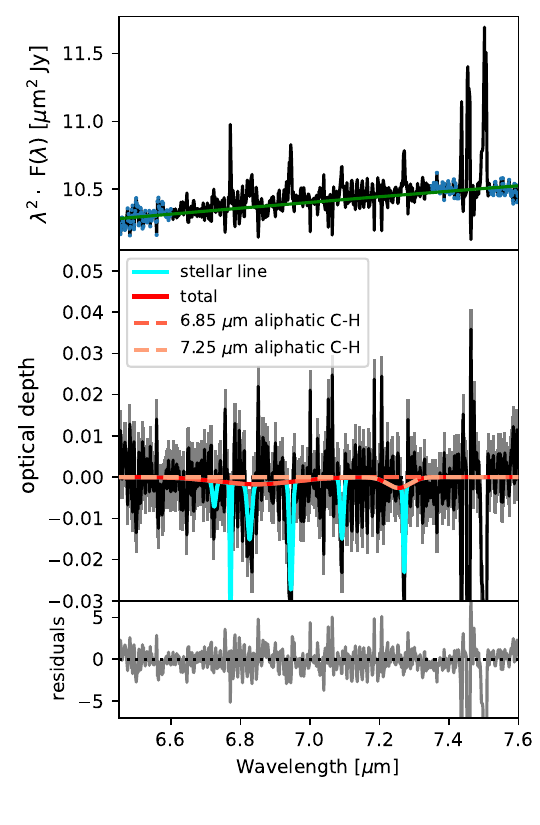}
    \caption{Here we show the MIRI MRS spectrum around 6.85~\micron\ and 7.25~\micron\ for GSC 08152-02121 (left) and CPD-59 5831 (right). In the top panel we show the observed spectrum in black, blue dots denoting data points from which the local continuum was calculated, and the fitted continuum as a green line. Between 7.43 and 7.53~\micron\ we encounter several stellar lines that are excluded from the continuum fit. The middle panel shows Gaussian components of the model fit, in units of optical depth, as dashed orange lines with the total as a red solid line. Stellar lines contaminating the fit are denoted by the cyan solid lines. The gray contours behind the data indicate the $1\sigma$ statistical uncertainty derived from the continuum. The bottom panel shows the significance of the residuals ($\pm5\sigma$) after subtraction of the continuum and line model, and normalization by the standard deviation.}
    \label{fig:c685725_absorption}
\end{figure*}

We detect several carbonaceous dust features in the \textit{JWST} spectra of GSC 08152-02121 and CPD-59 5831. The detected features are listed in Table \ref{tab:carbonir_features} and shown in Figs. \ref{fig:c34_absorption}, \ref{fig:c62_absorption}, and \ref{fig:c685725_absorption}.
We show representative local continuum fits and the data points used in the estimation in the top panels of Figs. \ref{fig:c34_absorption}, \ref{fig:c62_absorption}, and \ref{fig:c685725_absorption}. 
The middle panels of these figures show the fits to the carbonaceous dust features and any stellar lines within the regions of interest, whilst the bottom panels show the significance of the residuals (i.e. (observation - model)/uncertainty).
The uncertainties in the parameters of each dust feature have been derived in two ways. 
Firstly, each MCMC sampling for each continuum choice provides us with a statistical uncertainty. We present the average of these numbers as the statistical uncertainty. 
Secondly, from each of the MCMC samplings for each choice of continuum we collect the median value. We average these values to present as the mean feature strength and determine the standard deviation on these values as a measure of the systematic uncertainty caused by the choice of continuum. 
Ideally this value is substantially smaller than the statistical uncertainty derived from the MCMC fit itself. In practice, for our dataset, the value is of similar size as the statistical uncertainty. Both the statistical and systematic uncertainties are listed.
 \newline

\textbf{Carbonaceous dust features around 3.4~\micron:}
A set of four different local continua was defined to investigate the features around 3.4~\micron:
\begin{itemize}
    \item $2.5 - 2.8~\micron$ and $3.22-3.23~\micron$ and $3.60-3.8~\micron$ (maximum)
    \item $2.6 - 2.8~\micron$ and $3.22-3.23~\micron$ and $3.60 - 3.75~\micron$
    \item $2.7 - 2.8~\micron$ and $3.22-3.23~\micron$ and $3.60 - 3.70~\micron$
    \item $3.22 - 3.23~\micron$ and $3.60 - 3.70~\micron$ (minimum)
\end{itemize}
We do not include data between 2.8 and 3.22~\micron, because that wavelength interval may contain absorption due to water (see Section~\ref{sec:trapped_waterice}). 
The local continuum starts at 2.5~\micron, 
which is close to the starting point of our NIRCam spectra. At the long wavelength end we do not include data beyond 3.8~\micron\ to avoid a possible discontinuity between the NIRCam F332W and F444W LW filters near 4.0~\micron.
The carbonaceous dust features around 3.4~\micron\ were analyzed 
following the method described in~\citet{Chiar13}. We use five Gaussian models to decompose the aromatic and aliphatic absorption bands, representing the sp$^2$
CH, sp$^3$ $\mathrm{CH}_3$ and $\mathrm{CH}_2$ asymmetric, and sp$^3$ $\mathrm{CH}_3$ and $\mathrm{CH}_2$ symmetric modes. During the fitting process we fixed the standard deviation and mean of the Gaussians to the values from~\cite{Chiar13}, resulting in a fixed full-width at half-maximum (FWHM) and a fixed central wavelength. We left the amplitudes of the Gaussians free to vary. 
We explore the parameter space of the model using MCMC. We produce 300\,000 realizations of the model ($10\times n_{\mathrm{dim}} = 60$ walkers, 5\,000 steps), from which the posterior probability distribution is determined from the final 270\,000 realizations, where $n_{\mathrm{dim}}=6$, since we fit 5 carbonaceous features and one stellar line. Each realization of the model is compared to the observations using an error-weighted least squares as the log probability for that model. The posterior probability distributions are all monomodal and normally distributed. The maximum probability model was thus taken as the 50th percentile value of each parameter in the posterior. Uncertainties on the model parameters were determined by taking the difference between the 50th and 16th and 84th percentiles of the posterior as the $\pm$1-$\sigma$ uncertainties. 
\newline
\textbf{Carbonaceous dust features around 6.2~\micron:}
A set of four different local continua was defined to investigate the features around 6.2~\micron:
\begin{itemize}
    \item $5.60 -  5.75~\micron$ and $5.85 - 6.10~\micron$ and $6.42 - 6.60~\micron$ (maximum)
    \item $5.70 - 5.75~\micron$ and $5.85-6.10~\micron$ and $6.42-6.55~\micron$
    \item $6.0 - 6.1 ~\micron$ and $6.42 - 6.50~\micron$
    \item $6.05 - 6.1~\micron$ and $6.42 - 6.45~\micron$ (minimum)
\end{itemize}
The local continua do not include data between 5.75 and 5.85~\micron, because of the presence of an unidentified absorption feature (see Section~\ref{sec:trapped_waterice}). At the long wavelength end we include data up to 6.55~\micron\ to avoid dust features near 6.85 and 7.25~\micron (see below).

In the analysis of the $6.19$ and $6.25$~\micron\ features we fixed the central wavelength of the $6.25$~\micron\ feature with the value taken from~\citet{Chiar13}, while varying the remaining parameters. We produce 350\,000 realizations of the model ($10\times n_{\mathrm{dim}} = 70$ walkers, 5\,000 steps), from which the posterior probability distribution is determined from the final 315\,000 realizations. The model fit includes a stellar hydrogen line at $6.292~$\micron. \newline

\textbf{Carbonaceous dust features around 6.85~\micron\ and 7.25~\micron:} 
Again a set of four different local continua was defined:
\begin{itemize}
    \item $6.45 - 6.60~\micron$ and $7.35 - 7.43~\micron$ and $7.53 - 7.60~\micron$ (maximum)
    \item $6.51 - 6.60~\micron$ and $7.53 - 7.55~\micron$
    \item $6.52 - 6.60~\micron$ and $7.35 - 7.40~\micron$ 
    \item $6.52 - 6.60~\micron$ and $7.35 – 7.38~\micron$ (minimum)
\end{itemize}
At the short wavelength side we are limited by the 6.2~\micron\ feature. At the long wavelength end we include data up to 7.60~\micron\ to avoid changes in the slope arising from the broad silicate feature (see Section~\ref{sec:silicate}). The local continua do not include data between 7.43 and 7.53~\micron, because of the presence of several overlapping stellar lines that are difficult to fit together. 
We also follow the same strategy as for the 3.4~\micron\ features with values taken from~\citet{Chiar00}. The features are very weak in the spectra of both stars. For GSC 08152-02121 there may also be unidentified stellar lines present. Therefore, we only obtain upper limits for these features. 

\subsubsection{Silicate dust features}
\label{sec:silicate}
\begin{figure*}
    \centering
    \includegraphics[width=0.48\textwidth]{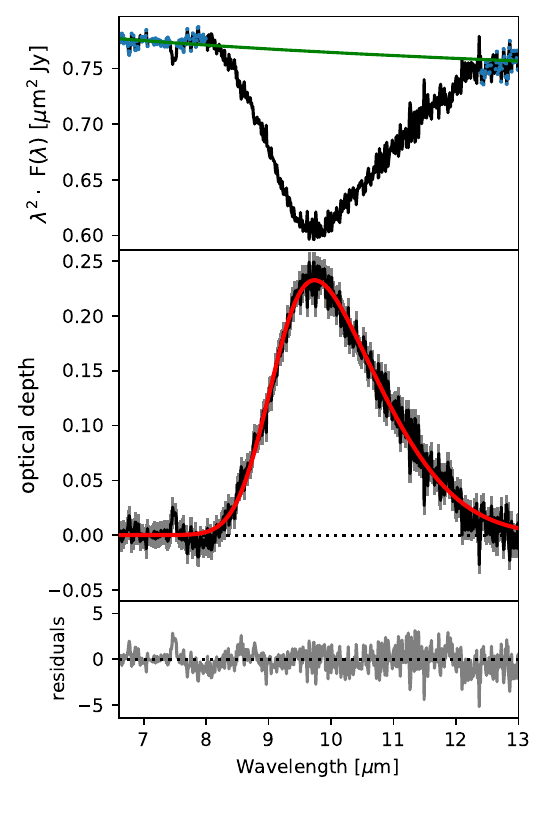}
    \includegraphics[width=0.48\textwidth]{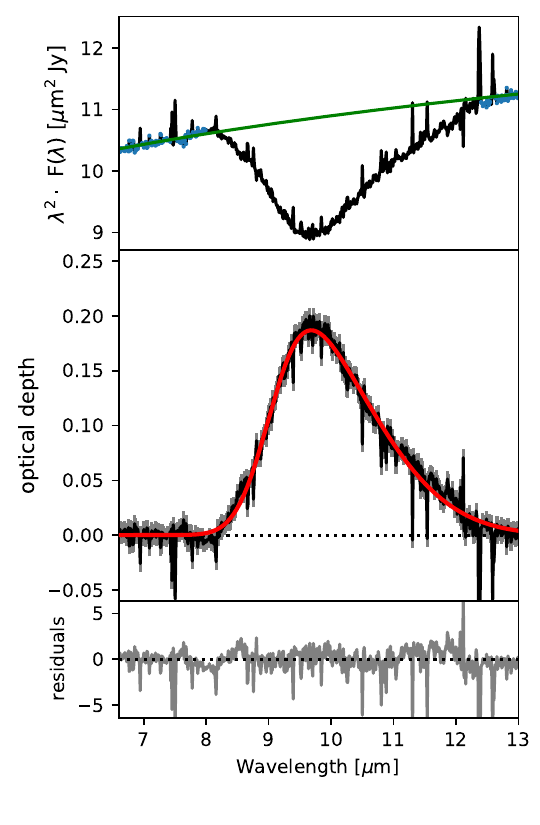}
    \caption{Here we show the MIRI MRS spectrum around the 10~\micron\ silicate absorption feature for GSC 08152-02121 (left), and CPD-59 5831 (right). For each target the top panel shows the observed spectrum in black, blue dots denote data points from which the continuum was calculated, and the fitted continuum as a green line. The stellar lines near 7.5~\micron\ are not included in the continuum fits. The middle panel shows a skewed Gaussian fit to the absorption as a red line, but it is inverted in units of optical depth; the continuum is shown as a dotted line. The gray contours behind the data indicate the $1\sigma$ statistical uncertainty derived from the continuum. The bottom panel shows the significance of the residuals ($\pm2.5\sigma$) after subtraction of the continuum and line model, and normalization by the standard deviation.}
    \label{fig:si_absorption}
\end{figure*}

The silicate absorption feature around 10~\micron\ is prominent in the MIRI MRS spectra of both targets. The feature at 20~\micron\ is only recovered on its short-wavelength side due to declining stellar flux and increasing noise from the telescope self-emission decreasing the spectral signal-to-noise beyond 15~\micron. Therefore, we omit the 20~\micron\ feature from the silicate feature fitting analysis. 
We identify regions of the MRS spectra on either side of the absorption feature, i.e. below 8~\micron\ and above 13~\micron, as being `continuum'. The exact slope and value of the continuum line was determined through a local least-squares fitting to the spectra. At these wavelengths broadband dust extinction reduces the observed flux below the stellar continuum \citep[see e.g.][]{VanBreemen2011}.
The selection of the continuum shape and wavelengths to which it is scaled therefore affects the resulting shape and depth of the absorption feature. A sloped line offers an acceptable representation of the out-of-feature extinction and stellar continuum, so we have adopted it here. The 10~\micron\ silicate absorption features for both targets are presented in Fig. \ref{fig:si_absorption}. After considering feature shapes such as a Gaussian or a (modified) Drude profile, we find that the feature is well represented by a skewed Gaussian profile, Equation 2.3 in \citet{Azzalini_2013}: 

\begin{equation}
f(x;A,\mu,\sigma,\gamma) = \frac{A}{\sigma\sqrt{2\pi}}e^{[-(x-\mu)^2/2\sigma^2]} \biggl\{ 1+\mathrm{erf}[\frac{\gamma(x-\mu)}{\sigma\sqrt{2}}]\biggl\}
\end{equation}

This model has four parameters: amplitude $A$, center $\mu$, sigma $\sigma$, and skew $\gamma$. We are interested in the maximum optical depth of the feature and the wavelength at which this occurs. 
Therefore, we need to calculate the mode, which can be approximated by the following equations:  \begin{equation}
 \mathrm{mode} \approx \mu + \sigma\,m_0 \mathrm{,\ with}
\end{equation}

\begin{equation}
 m_0 \approx \sqrt{\frac{2}{\pi}}\delta - \left( \left(1 - \frac{\pi}{4}\right)\frac{\left(\sqrt{\frac{2}{\pi}}\delta\right)^3}{1-\frac{2}{\pi}\delta^2}-\frac{\mathrm{sgn}(\gamma)}{2}\exp^{-\frac{2\pi}{|\gamma|}} \right) \mathrm{and}
 \end{equation}
 \begin{equation}\delta=\frac{\gamma}{\sqrt{1+\gamma^2}}.
\end{equation}

Using the skewed Gaussian, we also recover the FWHM and integrated area for the feature.

We explore the parameter space of the model using MCMC in the same way as described for the carbonaceous dust features in Section \ref{sec:carbon_irfeatures}. We produce 40\,000 realizations of the model ($10\times n_{\mathrm{dim}} = 40$ walkers with $n_{\mathrm{dim}}=4$, 1\,000 steps), from which the posterior probability distribution is determined from the final 36\,000 realizations. 
The results of the feature fitting are presented in Table \ref{tab:si_fit_results} and in Fig.~\ref{fig:si_absorption}.

We detect significant differences between the shapes of the silicate feature toward these two stars as has been seen previously \citep[][and references therein]{Gordon2021}. The larger integrated area for the GSC 08152-02121 feature is consistent with the higher source $A_{\rm V}$.

\begin{deluxetable*}{lccccc}
\tablecaption{Fitting results for the 10~\micron\ silicate feature. \label{tab:si_fit_results}}
\tablehead{\colhead{Source} & \colhead{$A_{\rm V}$ (mag)} & \colhead{Peak wavelength (Mode) (\micron)} & Optical depth &\colhead{FWHM (\micron)} & \colhead{Integrated Area ($\mathrm{cm}^{-1}$)}}
\startdata
%10 Lac & $0.210$ & $9.608\pm0.039$ & $0.016\pm0.001$ & $2.00\pm0.05$ & $\ 3.453\pm 0.075$ \\
GSC 08152-02121 & $5.100$ & $9.739\pm0.004$ & $0.233\pm0.001$ & $2.024\pm0.004$ & $49.835\pm 0.111$ \\ %2MASSJ08574757-4609145
CPD-59 5831 & $4.700$ & $9.688\pm0.004$ &  $0.187\pm0.001$ & $1.958\pm0.004$ & $39.267\pm 0.087$\\ %2MASSJ15095841-5958463
\enddata
\end{deluxetable*}

\subsection{Trapped water and other oxygen features}
\label{sec:trapped_waterice}
Shortward of the 3.4~\micron\ carbonaceous dust features, we encounter a feature that may be attributed to trapped water (see Fig.~\ref{fig:c34_absorption}). We clearly observe this feature in the spectrum of GSC 08152-02121, whereas it is absent in the spectrum of CPD-59 5831. The maximum optical depth of the feature toward GSC 08152-02121 can be derived from Fig.~\ref{fig:c34_absorption} to be 0.025. 
However, if this were associated with water, we do not see a commensurate absorption around 6.0 $\micron$ which would be associated with the O-H bending mode. We note that this feature may be shallow and difficult to detect, meaning its non-detection here does not categorically rule out the 3.0~$\micron$ feature being water related. 

We also report an absorption feature around 5.8~\micron (see Figure~\ref{fig:c62_absorption}). The origin of this feature is unclear, because a similar feature is present in the spectrum of the calibration star (stellar type O9~V) 10 Lac~\citep{Law2025}. In the case that the source of this absorption is astrophysical rather than instrumental, it may be associated with carbonyl (C=O) groups on interstellar dust grains.

\section{Discussion and future projects within WISCI}
\label{sec:dis}

%To borrow a phrase from the exoplanet community, to know the dust we must first know the star.
A detailed knowledge of the stellar atmosphere is essential to understand which features belong to the star in order not to confuse them with those imparted by interstellar dust. 
The spectral types of seven of the WISCI sample's stars were not precisely known before the program began. Amongst them were GSC 08152-02121 and CPD-59 5831 which were classified as OB- and OB+, respectively. The high resolution, high signal-to-noise VLT spectra enabled us to classify the stellar types of these two stars as B2IV and B5 Ia, respectively. 
 
We do not detect infrared excess in the spectra of GSC 08152-02121 and CPD-59 5831, which shows that neither of these stars has a strong stellar wind. However, winds also influence the strength of the hydrogen lines. The infrared contains many more hydrogen recombination lines than the UV and optical. Infrared lines with varying lower levels, e.g. Brackett and Pfund series, probe different regions in the wind than the Balmer series at shorter wavelengths \citep{2011Najarro}. 
This work contains an initial approach to match our spectra with new Tlusty models (Hubeny et al. submitted). In future work we will refine our stellar atmosphere model enabling us to analyse the dust features simultaneously and self-consistently over a broad wavelength range. The adoption of a locally determined stellar continuum to investigate absorption features already provides interesting insights into the dust along the lines of sight toward GSC 08152-02121 and CPD-59 5831. 

%Strong silicate feature detection, shape and peak are sightline dependent 
\textbf{Silicates:} We detect a strong silicate absorption feature in both GSC 08152-02121 and CPD-59 5831, as predicted by our modeling in Section~\ref{sec:feasibility}.
The shape of the 10~\micron\ silicate feature depends on the stoichiometry, composition, and size of the grains. Sub-micron grains produce relatively similar 10~\micron\ feature shapes regardless of their specific properties \citep{2001Bouwman}. However, a change in stoichiometry can still shift the absorption peak position \citep[e.g.][]{2003aJaeger}. The composition can also cause slight differences in the shape of the feature. In the case of an olivine type grain, the peak position of the feature is close to 10~\micron, but for a pyroxene stoichiometry this value is close to $9.4$~\micron~\citep{Fogerty16}. Amorphous silicate with a $\mathrm{Mg}_{1.5}\mathrm{SiO}_{3.5}$ composition from \citet{2003bJaeger}, produces a silicate peak at $9.6$~\micron. 
If we compare these values to the peak positions inferred from fitting, we find both sight lines have peak absorption values more consistent with amorphous silicate than pyroxene composition. We note that the different peak wavelengths may indicate further subtle differences between the sight lines in the stoichiometry and composition of the grains, or between their size and shapes \citep{2007Min}. However, this will require a more detailed analysis using silicate models based on laboratory spectra. 
In future research we will investigate all the sources in the WISCI program.
We will use laboratory data from e.g the Database of Optical Constants for Cosmic Dust of the Laboratory Astrophysics Group of the AIU Jena and SSHADE\footnote{\href{https://www.astro.uni-jena.de/Laboratory/OCDB/}{https://www.astro.uni-jena.de/Laboratory/OCDB/}, \href{https://www.sshade.eu/db/doccd}{https://www.sshade.eu/db/doccd}} in combination with new models of nanosilicates. 
We also will investigate possible substructure under the long wavelength side of the absorption feature. These substructures may indicate the presence of crystalline silicates and the detection of these features may give us information about crystallization processes and possible conversion mechanisms from the amorphous into the crystalline state and vice versa~\citep[e.g. laboratory studies of annealing silicate grains:][]{Nuth83,Hallenbeck1998,Hallenbeck2000}. These mechanisms are very important for understanding the processing of silicates in the interstellar medium~\citep{2010Henning}. Substructures in the silicate feature may also be produced by nanosilicates~\citep[e.g.,][]{Escatllar_2019, 2023Zeegers, Bromley2024}. 

\textbf{Carbonaceous dust:}
We detect the aromatic C-H, aliphatic C-H and C=C features at 3.3, 3.4 and 6.2 $\micron$ in the target spectra. The two sources show differences in their relative strengths. GSC 08152-02121 shows stronger absorption at 3.4~\micron\ than CPD-59 5831, whereas the 6.2~\micron\ features are stronger in CPD-59 5831. 
The extinction along the line of sight toward GSC 08152-02121 is higher than that of CPD-59 5831, indicating the presence of a larger amount of dust along this sightline. 
The dust may be distributed over several diffuse clouds or a single or a single diffuse cloud with a higher column density. Along the line of sight toward GSC 08152-02121 we may encounter a cloud with a higher column density than is typical. In this denser environment the grains may be more exposed to hydrogen, which may explain the differences in optical depth of the 3.4~\micron\ and 6.2~\micron\ features.  
For GSC 08152-02121 we find 3-$\sigma$ upper limits of 0.005 and 0.002 to the CH$_{2}$ and CH$_{3}$ bending features at 6.85 and $7.25$~\micron\, respectively.
The optical depths of these two features are large compared to measurements toward Sgr A*~\citep{Chiar00}. By taking into account the ratios found between the 3.4 and 6.85~\micron\ feature in \citet{Chiar00} and the optical depth of the 3.4~\micron\ feature found in this study, we would expect optical depths of $\sim 0.001$ for the 6.85~\micron\ feature. We note that the depth of features in GSC 08152-02121 may be over-estimated due to unidentified stellar lines. For CPD-59 5831, we do not detect any feature in most of the fits at this wavelength range. We obtain 3-$\sigma$ upper limits of 0.001 and 0.002 to the CH$_{2}$ and CH$_{3}$ bending features at 6.85 and 7.25~\micron, respectively.    
The WISCI project was underpinned by development of a novel code for modeling predictions of ISM absorption features, Section~\ref{sec:feasibility}. Here we predicted an optical depth of the strongest carbonaceous features of 1-2\% relative to the stellar continuum, which is indeed measured for the strongest features around 3.4~\micron. 

\textbf{Trapped water:} 
Along the line of sight toward GSC 08152-02121 we detect an absorption feature between $\sim 2.75 - 3.2~$\micron\, that may be related to the O-H stretching feature of water, as can be seen in Fig.~\ref{fig:c34_absorption}. This feature shows similarities to the feature detected by~\citet{Potapov21} in the spectrum of the bright reddened star Cyg OB 2 No. 12.
Their laboratory spectra of trapped solid-state water in silicate grains show a broad absorption from 2.75-3.2~\micron. The similar optical depth is surprising since the line of sight extinction toward GSC 08152-02121 is about half of that of Cyg OB 2 No. 12. In Section~\ref{sec:feasibility} we predicted lower optical depth feature of 0.4-0.8\% relative to the stellar continuum, which requires further investigation.
We do not detect any clear O-H-O bending feature near 6~\micron\ in this source. This feature is even shallower than the feature around 3~\micron\ and may not be detectable using the local stellar continuum fits used for this analysis. Instead, it requires a careful analysis using a detailed continuum extinction model and careful fitting of the stellar lines that may influence the shape of the feature. 
The origin and circumstances of this broad 2.75–3.2 $\micron$ feature which we associate with water is thus not immediately clear from Fig.~\ref{fig:c34_absorption} and its definitive assignment must await additional insights to be provided by analysis of the complete WISCI sample including all twelve sources.  
We see no comparable absorption in the same wavelength ranges for the CPD-59 5831 spectrum.

\textbf{How diffuse is the diffuse interstellar medium?} 
The two sight lines studied here are dissimilar, indicating the cloud densities and dust properties along these lines of sight are different. We note that the extinction per kpc is above the average of 2 magnitudes of extinction per kpc in the diffuse ISM toward GSC (2.38 $A_\mathrm{V}$ per kpc) and CPD-59 5831 (2.24 $A_\mathrm{V}$ per kpc), albeit by a lower amount in case of CPD-59 5831. 
We may thus expect a single denser cloud or multiple clouds contributing along the line of sight toward GSC. This may explain the stronger aliphatic carbonaceous dust features in the spectrum of this source and the possible presence of the trapped water feature. The detection of the C-H features indicates that these environments are not similar to the dense interstellar medium, where the 3.4\micron\ band is not seen. This is likely due to ices that form on a grain mantle once the material is dense enough to be shielded from the radiation field~\citep{Mennella_02_CH}. Although the C-H bond is destroyed in the diffuse ISM by interaction with UV radiation, it can also be reformed~\citep{Mennella_02_CH}, which may explain why we detect the features in our spectra. 
We note that some of the dust features may only be present in the densest regions along the lines of sight, while other components are more likely found in the lower density regions.
Further investigation of the remaining WISCI sample is necessary to identify trends for these features in the diffuse ISM. 

\textbf{Future projects within WISCI:}

Future work will focus on a detailed analysis of the stellar properties for all the sources in the program.
Combining the new UV, optical, and IR spectra gathered for WISCI with stellar atmosphere models will provide an in-depth characterization of the stellar physical parameters and photospheric emission, critical to assessment of the astrophysical nature of absorption features seen in their spectra. The WISCI data also enables a detailed study of the structure of stellar winds and mass loss-rates~\citep[e.g. ][ and future work]{2024Law}. 
Future work will also include measuring the UV-MIR extinction curve including continuum and feature extinction, which will provide high fidelity feature profiles and full extinction measurements for inclusion in dust grain models. We will study the silicate, water and carbonaceous dust features along all sight lines. There we will also provide a comparison of the full WISCI sample with the MEAD~
\citep{Decleir2025} and Ice Age~\citep{2023McClure} studies. 
We will analyze the infrared-detected features together with dust features at shorter wavelengths for each source: the DIBs, 2175 \AA{} feature, intermediate scale structures and the overall extinction (Gordon et. al in prep and Wei et al. in prep.). 

\section{Conclusions}
\label{sec:con}

We presented the motivation and project summary for the WISCI project combining data from ultraviolet to mid-infrared wavelengths to trace the absorption of dust in the diffuse interstellar medium toward twelve stars. In this work we presented observations of two stars in the sample as a `first shot' at demonstrating the potential of the WISCI program to reveal the composition and structure of dust grains in the interstellar medium.

We have measured the stellar photospheres of GSC 08152-02121 and CPD-59 5831 from UV to mid-IR wavelengths for the first time. The stellar spectra show characteristic reddening due to interstellar extinction and exhibit strong emission lines out to mid-infrared wavelengths.

We detect silicate and carbonaceous dust features from interstellar dust in the spectra of both stars. We find the shape, central wavelength, and relative depth of both species to be sightline dependent. The silicate feature of GSC 08152-02121 has a longer central wavelength, and is deeper and broader than that of CPD-59 5831. This correlates with overall extinction, $A_{\rm V}$, being higher toward GSC 08152-02121. 
The carbonaceous dust feature at 3.3~\micron\ is weaker for GSC 08152-02121 whereas the adjacent 3.4~\micron\ feature is stronger for GSC 08152-02121. At 6.19 and 6.25~\micron\ both features are stronger for CPD-59 5831, whilst at 6.85 and 7.25~\micron\ the upper limits on the features once again indicate possible stronger features in the GSC 08152-02121 spectrum than in CPD-59 5831.
This difference in behavior may be attributed to the origins of the 3.3 and 3.4~\micron\ features originating from C-H bonds, whereas the 6.19 and 6.25~\micron\ features arise from C=C bonds. These bonds may persist in spectra toward sight lines tracing diffuse regions, since they are less prone to photodissociation. 

In addition to carbonaceous and silicate features, we detect tentative evidence of solid-state water (interpreted as hydrated silicates) at 3.0~\micron\ in the diffuse ISM toward GSC 08152-02121. These observations, after the detection along the line of sight of Cyg OB 2 No. 12 \citep{Potapov21}, provides the clearest observations of solid water in the diffuse ISM to date. It is not yet understood why the feature would be as deep along this sightline as toward Cyg OB 2 No. 12 because of their different extinction values, nor why CPD-59 5831 has no detectable water absorption despite its overall extinction being comparable to GSC 08152-02121.

Future work in the WISCI project will: 
\begin{itemize}
    \item Provide a detailed analysis of the stellar properties for all the sources in the program.
    \item Extend the extinction analysis to the IR.
    \item Contain detailed studies of the shape and structure of the dust grains, and detailed work to extract features associated with the presence of nanoparticles along the line of sight of all twelve sources.
    \item Identify trends between the strength of silicate, carbonaceous, and water features to the overall extinction.
\end{itemize}

\section{acknowledgements}
This research has made use of the following databases: NASA’s Astrophysics Data System, the SIMBAD database, 
operated at CDS, Strasbourg, France \citep{2000Wenger}.
This work is based [in part] on observations made with the NASA/ESA/CSA James Webb Space Telescope. The data were obtained from the Mikulski Archive for Space Telescopes at the Space Telescope Science Institute, which is operated by the Association of Universities for Research in Astronomy, Inc., under NASA contract NAS 5-03127 for \textit{JWST}. These observations are associated with program \#2183.
This research is based on observations made with the NASA/ESA Hubble Space Telescope obtained from the Space Telescope Science Institute, which is operated by the Association of Universities for Research in Astronomy, Inc., under NASA contract NAS 5–26555. These observations are associated with program \#17078.
The uncalibrated JWST and HST data presented in this article were obtained from MAST at the Space Telescope Science Institute. The uncalibrated JWST data files used in this work can be accessed from MAST via \dataset[doi: 10.17909/e5mt-cq07]{https://doi.org/10.17909/e5mt-cq07}.
Based on observations collected at the European Southern Observatory under ESO programmes 109.23GA and 112.25JW.
Based on observations collected at the Himalayan Chandra Telescope under programmes HCT-2022-C2-P30 and HCT-2022-C3-P25.
All the reduced data sets used to analyze the dust features and stellar types of the stars GSC 08152-02121 and CPD-59 58319 can be obtained via Zenodo: 
\url{https://doi.org/10.5281/zenodo.15253941}.
This publication makes use of data products from the Wide-field Infrared Survey Explorer, which is a joint project of the University of California, Los Angeles, and the Jet Propulsion Laboratory/California Institute of Technology, funded by the National Aeronautics and Space Administration.
Figure~\ref{fig:map} dust map image credits: ESA/Gaia/DPAC - CC BY-SA 3.0 IGO. Acknowledgements: Created by T.E. Dharmawardena, \textit{Gaia} group @ MPIA
This work has made use of data from the European Space Agency (ESA) mission
 (\href{https://www.cosmos.esa.int/gaia}{\it{Gaia}}), processed by the {\it Gaia}
Data Processing and Analysis Consortium (\href{https://www.cosmos.esa.int/web/gaia/dpac/consortium}{DPAC}). Funding for the DPAC
has been provided by national institutions, in particular the institutions
participating in the {\it Gaia} Multilateral Agreement. 
Two workshops were organized to discuss the observations of the WISCI project. We thank the Lorentz Center for supporting the workshop ``Interstellar Soot and Sand with JWST'' in April 2023. The Lorentz Center is funded by the Dutch Research Council (NWO) and the University of Leiden. The second workshop took place at STScI in Baltimore. 
We acknowledge support from ESA through the Science Faculty - Funding reference ESA-SCI-E-LE-005. 
We acknowledge travel support from STScI - Funding reference TER-TA24-03-0460F. % add travel funding Jonty and Burcu. 
SZ and MD acknowledge support from the Research Fellowship Program of the European Space Agency (ESA).
JPM acknowledges support by the National Science and Technology Council of Taiwan under grant NSTC 112-2112-M-001-032-MY3.
BG acknowledges support of T\"UB\.ITAK 2219 program.
AP acknowledges support from the German Federal Ministry for  Economic Affairs and Climate Action (the German Aerospace Center project 50OR2215) and from the Deutsche Forschungsgemeinschaft (Heisenberg grant PO 1542/7-1). TRG's research is supported by the international Gemini Observatory, a program of NSF NOIRLab, which is managed by the Association of Universities for Research in Astronomy (AURA) under a cooperative agreement with the U.S. National Science Foundation, on behalf of the Gemini partnership of Argentina, Brazil, Canada, Chile, the Republic of Korea, and theUnited States of America. 
This project has received funding from the European Research Council (ERC) under the European Union's Horizon 2020 research and innovation programme (grant agreement 101164755/METAL) and  from the Israel Science Foundation (ISF) under grant number 2434/24.
AP acknowledges support from the German Federal Ministry for  Economic Affairs and Climate Action (the German Aerospace Center project 50OR2215) and  from the  Deutsche Forschungsgemeinschaft (Heisenberg grant PO 1542/7-1).
This work was partly supported by the Spanish program Unidad de Excelencia Mar\'ia de Maeztu CEX2020-001058-M, financed by MCIN/AEI/10.13039/501100011033.
We thank the staff of IAO, Hanle and CREST, Hosakote, that made the HCT observations possible. The facilities at IAO and CREST are operated by the Indian Institute of Astrophysics, Bangalore.
\vspace{5mm}
\facilities{\textit{HST} STIS \citep{STIS}, VLT XSHOOTER \citep{XSHOOTER}, \textit{JWST} NIRCam and MIRI \citep{JWST,NIRCam,MIRI,MIRIMRS}. HCT HESP and HFOSC}
\vspace{5mm}
\software{In addition to the packages mentioned in the text, this work has made use of the following Python modules: Astropy \citep{astropy2013,astropy2018,astropy2022}, Corner \citep{2016ForemanMackey}, Emcee \citep{2013ForemanMackey}, Matplotlib \citep{Matplotlib}, Numpy \citep{NumPy}, Scipy \citep{SciPy}, and SynPhot \citep{Synphot}, and the \href{https://asd.gsfc.nasa.gov/archive/idlastro/}{IDL Astronomy User's Library}}

\appendix

\section{Summary of WISCI observations and data reduction}
\label{app:obs_summary}

The Appendix contains a detailed description of the observations. An overview of all the observations within the WISCI program can be found in Table~\ref{tab:obs_summary}. 

\begin{table}[!h]
\caption{Summary of WISCI observing programs. Observations were taken with \textit{JWST} NIRCam and MIRI MRS, \textit{HST} STIS, VLT XSHOOTER, and HCT HESP or HFOSC. \label{tab:obs_summary}}
\centering
\def\arraystretch{1.1}
 \begin{tabular}{l c c c c c c c c } 
 \hline
 2MASS J[...] & \textit{JWST} & PID  &   \textit{HST}   & PID  &      VLT    & PID  & HCT & PID \\
              &   &  &   &  &     & &   &  (HCT-2022-) \\ 
 \hline08574757$-$4609145  & \checkmark & 2183 & \checkmark & 17078 & \checkmark & 112.25JW & \\ 
 13015278$-$6131045  & \checkmark & 2183 & \checkmark & 17078 & \checkmark & 109.23GA & \\ 
 15095841$-$5958463  & \checkmark & 2183 & \checkmark & 17078 & \checkmark & 109.23GA & \\ 
 17075654$-$4040383  & \checkmark & 2183 &  \checkmark & 17078 & \checkmark & 112.25JW & \\ 
 17362876$-$3253166  & \checkmark & 2183 &  \checkmark & 17078 & \checkmark & 112.25JW & \\   
 18112983$-$2017075  & \checkmark & 2183 &  \checkmark & 17078 & \checkmark & 109.23GA &\checkmark  & C2-P30 \\   
 18230252$-$1320387  & \checkmark & 2183 &  \checkmark & 17078 & \checkmark & 112.25JW & \checkmark  & C2-P30, C3-P25 \\  
 20311055+4131535  & \checkmark & 2183 &  \checkmark   & 17078 & &  & \checkmark & C2-P30 \\ 
 20323486+4056174  & \checkmark & 2183 &  \checkmark   & 17078 & &  & \checkmark & C2-P30, C3-P25 \\  
 20331106+4110321  & \checkmark & 2183 &  \checkmark   & 17078 & &  & \checkmark & C2-P30, C3-P25 \\ 
 20332674+4110595  & \checkmark & 2183 &  \checkmark   & 17078 & &  & \checkmark & C2-P30, C3-P25 \\ 
 20452110+4223513  & \checkmark & 2183 &  \checkmark   & 17078 & &  & \checkmark & C2-P30, C3-P25 \\ 
 \hline 
\end{tabular}
\end{table}

\subsection{JWST NIRCam}
As the sources in our sample exceed NIRSpec brightness limits, the only option for obtaining spectra below 5~\micron\ 
was to utilize the NIRCam GRISM.  
Our goal for these observations was a S/N = 300 at a wavelength of 3.4~\micron\ and spectral resolution of 0.01~\micron\ ($R=1500$). 
To reach this criterion we average the observed spectrum over 9.8 pixels to obtain the required S/N.
NIRCam GRISM observations in Cycle 1 were supported using two templates in 
APT\footnote{\url{https://www.stsci.edu/scientific-community/software/astronomers-proposal-tool-apt}}, 
wide field slitless spectroscopy (WFSS) and grism time series (TSO)~\citep{Greene2017}.  
With WFSS, observations are restricted to full frame readouts and the corresponding single frame time ($\sim$10.73~sec) 
exceeded saturation limits for these targets. 
Therefore, we employed the grism TSO template which supports three subarrays, SUBGRISM64, SUBGRISM128, and 
SUBGRISM256, allowing per frame exposure times of $\sim$0.34, 0.68, and 1.35~sec, respectively when read 
through 4 detector outputs.
Depending on source brightness and GRISM, 25\% of our data were obtained in SUBGRISM128 and the remainder in SUBGRISM256.
The selection of subarray was driven by ensuring sufficient samples up the ramp 
before saturation to provide some mitigation against data loss from cosmic rays for a given source
brightness, resulting in 4 integrations, each with 5 groups. 
All objects were observed using GRISM-R (dispersion in the row - fast read - direction) and both the 
F322W2 and F444W filters, resulting in spectral coverage from $\sim$2.4 to 4~\micron\ (F322W2) and 
$\sim$3.9 to 5~\micron\, (F444W) at a resolution between 1200 and 1600. 
In the grism TSO mode, the simultaneously observed SW arm of NIRCam is programmatically restricted to 
using weak lens `filters'; we utilized the WLP8 weak lens which introduces 8 wavelengths of positive 
defocus at the focal plane.

As the reduction for these data paralleled that implemented for PID-02459 (``MEAD''), we highlight the approach here and point the interested reader to the more detailed summary of the process given in \citet{Decleir2025}. 

\begin{itemize}
    \item Data were reduced through level 2 using custom software developed during instrument design and testing.
    \item An empirical PSF was constructed at the TSO field point, using program data, PID-02459 data and appropriate
    commissioning data (PID-01076).  The spectral trace from \citet{2023Sun} was used with small per-object adjustments 
    where necessary. 
    \item The empirical PSF was used to extract a 1-D spectrum for  each program object in native (DN/sec) units. 
    \item A 1-D spectrum of the flux calibrator P330-E was extracted from data obtained at the TSO field point in 
    commissioning (PID-01076) in an identical fashion.
    \item The wavelength calibration was taken from \citet{2023Sun} and verified against program objects with 
    identifiable spectral features. 
    \item Data were flux calibrated by first dividing the extracted program spectra by the 1-D spectrum of 
    P330-E in pixel space before convolving with the latest P330-E CALSPEC SED in wavelength space
    \citep{2020AJ....160...21B}.  This was done to mitigate residual column noise in the final spectra that
    result from the lack of dithering in the TSO mode but are common to program and calibrator and so can 
    be divided out.
\end{itemize}

We note that for data in the F444W/GRISM, this approach to flux calibration can not be used as the 
TSO field point for the F444W/GRISM was changed post commissioning to better center the source spectrum
on the detector. Therefore, no calibrator data at the TSO field point were available and the normal approach of deriving a sensitivity curve and calibrating the program objects in wavelength space 
must be used. This results in higher detector residual systematic noise in our F444W spectra relative
to the F322W2 spectra. As part of the ongoing \textit{JWST} calibration program, data will be obtained in TSO
mode at the nominal field point, allowing us to revisit the F444W calibration in the future. 

\subsection{\textit{JWST} MIRI MRS}
The MIRI MRS observations were specified for a S/N = 300 and spectral resolution of 0.01~\micron\ at 10~\micron. To achieve this we aim for a S/N = 140 per pixel in the observed spectrum \cite[$R \simeq 3000$ at 10~\micron\ sampled by $\simeq$1.5~pixels,][]{Labiano21} and then average that over 4.5 pixels to achieve the required S/N. The targets were observed with a 4-point dither pattern suitable for point sources. We found a substantial decrease in the quality of the observations at longer wavelengths, specifically channel 4, consistent with the declining stellar flux, increasing self-noise from the telescope's thermal emission, and the known low throughput of this channel.

The spectra have been reduced using the \textit{JWST} pipeline (version 1.12.5) with calibration data system pmap 1150 reference files.
Overall the standard reduction was done with a few modifications.
In the second stage of the pipeline (CALWEBB\_SPEC2) the residual fringe correction step was skipped.
In the third stage of the pipeline (CALWEBB\_SPEC3), the outlier detection in the cube building step was done and in the spectral extraction step the autocentering and residual fringing was used.
A post-processing step was done to remove the MRS spectral leak that shows up 
as a broad feature at $\sim$12.3 \micron\ \citep[see Sect. 2.5.6 of][]{Gasman2023}.
The twelve MIRI MRS orders were combined, along with the two NIRCam orders, into a single merged spectrum using the overlap between orders to perform a multiplicative correction.
There were a few noise spikes in the spectra likely due to residual cosmic ray hits and these were removed with a simple outlier detection algorithm. 
The results were visually inspected to ensure no emission/absorption lines were affected.

\subsection{\textit{HST} STIS}

We use \textit{HST} STIS (PID: 17078, PI: Zeegers) to observe 0.115 -- 1~\micron\ spectra of our targets at low spectral resolution ($\mathrm{R}\sim1000$).
We cover the 0.115 -- 0.30~\micron\ range using the G140L and G230L gratings with STIS/MAMA, while the optical (0.29 -- 1~\micron) range was observed using the G430L and G750L gratings with STIS/CCD. 
To minimise losses we use the 52X2 slit for all four gratings.
Since we are primarily interested in the continuum, we determine our target S/N at a resolving power of 10, aiming for S/N $> 10$ in the UV, and S/N $> 100$ in the optical. Shallow and broad dust features such as the intermediate scale structures~\citep{Massa20} greatly benefit from the photometric accuracy delivered by \textit{HST} STIS. The processing of the ground-based data tends to absorb these features into the continuum fit.
As we are targeting stars that are highly-extinguished at UV wavelengths, the flux at the shortest wavelengths is too low to achieve higher S/N in the UV.
The data were retrieved from the \textit{HST} archive in the final 1D extracted form
(i.e. x1d and sx1 files).

\subsection{VLT XSHOOTER}

Out of the twelve WISCI targets, seven are observable from the Southern hemisphere. These were observed by the XSHOOTER spectrograph \citep{XSHOOTER} on the VLT by programs 109.23GA and 112.25JW (PI: Zeegers). Program 109.23A was accepted as a filler program and was able to complete three targets, including CPD-59 5831. The remaining four targets, including GSC 08152-02121, were covered by program 112.25JW.
Throughout the programs the slit widths were set to maximize photometric accuracy while providing a spectral resolution in excess of $R=3200$ (1\farcs6, 1\farcs5, 1\farcs2 wide for UVB, VIS and NIR respectively). In practice the resolution was measured to be $R>4000$ in the spectral typing region between 4000 and 5000 \AA{}. For GSC 08152-02121 the exposure times were 400~sec, 463~sec and 500~sec, respectively. For CPD-59 5831 the exposure times were set to 30~sec, 93~sec and 160~sec, respectively. The exposure times of the VIS and NIR branch were increased to fill the dead-time of the detector readouts and therefore not equal in length to the UVB branch exposure. The \texttt{XSHOOTER\_slt\_obs\_Stare} template was used to reduce telescope overheads by not using nodding, as the NIR branch data was not seen as critical. In practice, the thermal background only starts contributing noticeably at wavelengths longer than 2.3~\micron.

Part of the spectrum of CPD-59 5831 across seven orders in the VIS branch ($5500-10200\, $\AA{}) was saturated as the seeing was substantially better (0\farcs63) than anticipated (requirement $<1\farcs5$) which reduced the slit losses. The critical stellar typing region between 4000 and 5000 \AA{} falls in the UVB branch and was therefore not impacted. This issue only occurred for this single target and its scientific impact was minimal.

{Data reduction:} We used the standard ESO XSHOOTER pipeline (v3.6.3) in combination with the EsoReflex workflow \citep{Freudling13} and default settings with enabled \texttt{starestack} to reduce the data of the three instrument arms. The \texttt{starestack} setting combines the multiple frames of the NIR data into one data product.

{Telluric correction:} Starting at the 1D extracted (\texttt{\_merge1d}) files output by the standard pipeline we applied \texttt{molecfit} \citep[v4.3.1;][]{2015Smette} to the 1D merged spectra. We use the \texttt{molecfit} settings and the wavelength regions of Tables 1 and 2 of \cite{2020Gonneau}, respectively, to fit the telluric absorption. For the NIR and VIS branches the wavelength regions of their Table 2 also match the regions used by \cite{2015Kausch}. Note that the telluric correction in the UVB branch only has a low order minor effect close to the transition with VIS, and a stronger effect on the bluest end of the spectrum where the data has a low S/N which is therefore excluded from our fits.

The telluric corrected 1D spectra were subsequently fitted with cubic splines following the convex-hull / alpha shape principle via the \texttt{RASSINE} package \citep{2020Cretignier}. Care was taken to prevent broad spectral features associated with the star or ISM (like DIBs) to be fitted with the continuum. However, as mentioned before, it is expected that extremely broad and shallow dust features such as the intermediate scale structures~\citep{Massa20} are fitted by the continuum. A barycentric correction was applied to the wavelength axis. 

\subsection{HCT HFOSC \& HESP}

The seven northernmost sources in the WISCI sample were observed using the Himalayan Chandra Telescope (HCT)\footnote{\url{https://www.iiap.res.in/centers/iao/facilities/hct/}}, equipped with both the Hanle Echelle Spectrograph (HESP, $3500-10000\ $\AA{}) and the Himalaya Faint Object Spectrograph (HFOSC, $3500-9000\ $\AA{}) using Grism 7 and Grism 8. Three sources LS 4992, TYC 6272-339-1, and VI Cyg 1  were observed with HESP at a high spectral resolution (of $R = 30\,000$), allowing for detailed spectroscopic analysis. However, the remaining four sources A38, ALS 15181, GSC 03157, and 2MASS J2045+4223 were too faint to be observed at such a high resolution with HESP. As a result, these sources were observed with the HFOSC instrument at a lower spectral resolution ($R = 1500-2500$), which was more suitable given their faintness. Flat frames were taken prior to each on-target observation, and bias frames were taken both before and after the on-target observations. Lamp spectra were obtained for wavelength calibration for both the HESP and HFOSC observations. Additionally, during the HFOSC observations, we observed spectral standard stars to correct the low-resolution spectra for instrumental/telluric effects.

The reduction of the high-resolution HESP data was performed using the HESP pipeline\footnote{\url{https://github.com/arunsuryaoffice/hesp_pipeline}}. This pipeline handles several preprocessing steps, including bias subtraction, overscan correction, trimming, and Cosmic Ray correction, before extracting the spectral orders. Once the orders are extracted, a wavelength solution is applied to the spectrum. The spectrum is then blaze-corrected and normalized, which results in the final extracted spectra not retaining the information on the shape of the underlying continuum.

The HFOSC data were reduced in a standard procedure with the HCT pipeline  Hfosc-Automated-PIpeLIne (HAPLI) \citep{2023Narang}, which performs dark, bias correction, and flat fielding and automatically does the aperture identification and extraction. This is followed by wavelength calibration to produce the final spectra. The final spectra where then corrected for instrumental/telluric effects by utilizing the spectral standard stars.

\bibliography{wisci-shot}{}
\bibliographystyle{aasjournal}



\end{document}